\pdfoutput=1 
\documentclass[english,structabstract]{aa}
\usepackage{mathptmx}
\usepackage[T1]{fontenc}
\usepackage[utf8]{inputenc}
\usepackage{graphicx}

\makeatletter
\bibpunct{(}{)}{;}{a}{}{,} 
\usepackage{txfonts}

\newcommand{\teff}{T_{\mathrm{eff}}}
\newcommand{\logg}{\log g}
\newcommand{\feh}{\left[\mathrm{Fe}/\mathrm{H}\right]}
\newcommand{\ltaur}{\log\tau_{\mathrm{Ross}}}
\newcommand{\taur}{\tau_{\mathrm{Ross}}}

\newcommand{\agran}{A_{\mathrm{gran}}}
\newcommand{\dgran}{d_{\mathrm{gran}}}

\titlerunning{The Stagger-grid -- VI. Surface appearance of stellar granulation}
\authorrunning{Z. Magic and M. Asplund}

\@ifundefined{showcaptionsetup}{}{%
 \PassOptionsToPackage{caption=false}{subfig}}
\usepackage{subfig}
\makeatother

\usepackage{babel}
\begin{document}

\title{The \textsc{Stagger}-grid: A grid of 3D stellar atmosphere models}

\subtitle{VI. Surface appearance of stellar granulation}

\author{Z. Magic\inst{1,2} \and  M. Asplund\inst{2}}

\institute{Max-Planck-Institut f{\"u}r Astrophysik, Karl-Schwarzschild-Str.
1, 85741 Garching, Germany\\
\email{magic@mpa-garching.mpg.de} \and  Research School of Astronomy
\& Astrophysics, Cotter Road, Weston ACT 2611, Australia}

\offprints{magic@mpa-garching.mpg.de}

\date{Received ...; Accepted...}

\abstract{In the surface layers of late-type stars, stellar convection is manifested
with its typical granulation pattern due to the presence of convective
motions. The resulting photospheric up- and downflows leave imprints
in the observed spectral line profiles.}{ We perform a careful statistical
analysis of stellar granulation and its properties for different stellar
parameters. }{We employ realistic 3D radiative hydrodynamic (RHD)
simulations of surface convection from the \textsc{Stagger}-grid,
a comprehensive grid of atmosphere models that covers a large parameter
space in terms of $\teff$, $\logg$, and $\feh$. Individual granules
are detected from the (bolometric) intensity maps at disk center with
an efficient granulation pattern recognition algorithm. From these
we derive their respective properties: diameter, fractal dimension
(area-perimeter relation), geometry, topology, variation of intensity,
temperature, density and velocity with granule size. Also, the correlation
of the physical properties at the optical surface are studied.}{
We find in all of our 3D RHD simulations stellar granulation patterns
imprinted, which are qualitatively similar to the solar case, despite
the large differences in stellar parameters. The granules exhibit
a large range in size, which can be divided into two groups -- smaller
and larger granules -- by the mean granule size. These are distinct
in their properties: smaller granules are regular shaped and dimmer,
while the larger ones are increasingly irregular and more complex
in their shapes and distribution in intensity contrast. This is reflected
in their fractal dimensions, which is close to unity for the smaller
granules, and close to two for larger granules, which is due to the
fragmentation of granules. }{Stellar surface convection seems to
operate scale-invariant over a large range in stellar parameters,
which translates into a self-similar stellar granulation pattern.}


\maketitle

\section{Introduction\label{sec:Introduction}}

\begin{figure*}
\hspace{17mm}$8.0\,\left[\mathrm{Mm}\right]$\hspace{33mm}$28\,\left[\mathrm{Mm}\right]$\hspace{32mm}$2400\,\left[\mathrm{Mm}\right]$
\hspace{30mm}$1.4\,\left[\mathrm{Mm}\right]$

\includegraphics[width=184mm]{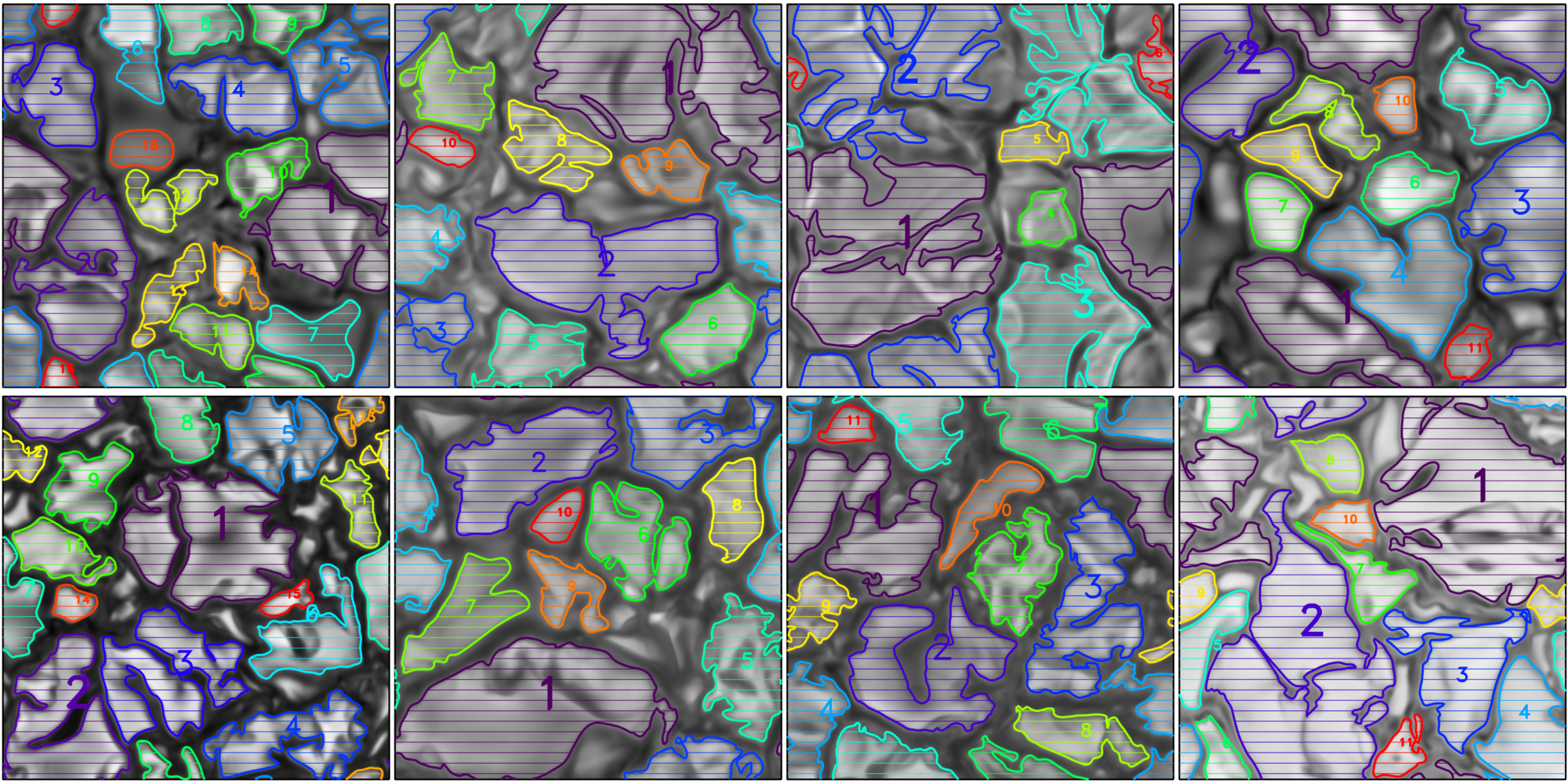}

\hspace{17mm}$7.2\,\left[\mathrm{Mm}\right]$\hspace{30mm}$22\,\left[\mathrm{Mm}\right]$\hspace{32mm}$1500\,\left[\mathrm{Mm}\right]$
\hspace{30mm}$1.2\,\left[\mathrm{Mm}\right]$\caption{\label{fig:gran_detection}Emergent (bolometric) intensity map (\emph{gray
contour}s) over-plotted with the colored-coded contours of the recognized
granules with solar (\emph{top panel}) and \emph{$\feh=-3.0$} (\emph{bottom}).
From \emph{left} to \emph{right}: the Sun ($\teff/\logg=5777\,\mathrm{K}/4.44$),
turnoff star ($6500\,\mathrm{K},\,4.0$), K-giant ($4500\,\mathrm{K},\,2.0$)
and K-dwarf ($4500\,\mathrm{K},\,5.0$). The granules are numbered
in order of decreasing granule size at its respective barycenter (large
characters refers to large granule size).}
\end{figure*}
In the envelopes of cool stars, the energy resulting from the nuclear
burning at the center is transported through convective energy transport,
which involves the ascension of hot, buoyant plasma towards the surface.
At the optical surface, the overturning convective motions into downdrafts
do not come to rest immediately, instead the upflows overshoot well
into the visible photosphere due to its inertia, hence leaving an
imprint in the emergent radiation in form of a typical granulation
pattern. The visible stellar surface of late-stars is patterned with
bright elements (granules) interspersed by the dark intergranular
lane. Understanding convection is an important aspect for the energy
transport in late-type stars, however, this is a non-trivial task
due to the non-linear and non-local nature of (turbulent) surface
convection.

The Sun shows a distinct granulation pattern on its observable (optical)
surface, which is the manifestation of convection that transports
energy to the surface. The solar granulation pattern has been subject
to manifold observational studies over the last decades with progressively
increasing resolution due to technological advances \citep[e.g.][]{Roudier:1986p20337,Hirzberger:1997p16292,Hirzberger:1999p16391,Hirzberger:1999p16220,Schrijver:1997p16291,Bovelet:2007p6139,Bovelet:2001p6138,Abramenko:2012p20140}.
Nowadays high-resolution solar observations are comparable to the
typical numerical resolution with a few tens of kilometers. The observational
developments were accompanied by improvements in the detection and
derivation of statistical properties of solar granulation. Moreover,
details of the solar small-scale magnetic structures has also been
studied \citep{Solanki1993SSRv...63....1S,Janen:2003p11810,Carlsson:2004p12217,Abramenko:2005p20173,Stein:2006p6127,Wiehr:2009p6135}.

Until the advent of realistic 3D radiative hydrodynamic (RHD) simulations,
which involves the (computationally expensive) solution of the hydrodynamic
equations coupled with a realistic radiative transfer \citep{Nordlund:1982p6697,Steffen:1989p18861},
direct comparisons of theoretical predications with the solar granulation
properties were absent. In contrast to theoretical 1D models, such
3D models are capable of predicting the typical stellar granulation
pattern imprinted in the (bolometric) intensity map emerging from
the stellar surface. They have revealed that stellar surface convection
is driven by the large-amplitude entropy fluctuations in a thin optical
surface boundary layer, where the energy can escape into space \citep[see][]{Stein:1998p3801,Nordlund:2009p4109}.
The dark intergranular lanes stem from the entropy-deficient plasma
that descends into narrow turbulent downdrafts, while the granules
are warm upflowing plasma. These flows exhibit a distinct asymmetry
in its thermodynamic properties, leading to inhomogeneities and velocities,
which has been compared with observations of the Sun \citep[e.g.,][]{Asplund:2000p20875,Asplund:2004p7805,Carlsson:2004p12217,Pereira:2009p17405,Pereira:2009p17397,Pereira:2013arXiv1304}
and other stars \citep[e.g.,][]{Nordlund:1990p6720,Collet:2006p6298,Collet:2007p5617,Ramirez2008A&A...492..841R,Ramirez:2009p10905,Ramirez2010ApJ...725L.223R,Chiavassa:2009p22491,Chiavassa:2010p22497,Chiavassa:2011p22500,Chiavassa:2012p22493}.
These comparisons have confirmed the realism of the 3D RHD simulations.
Due to the increasing computational power, various grids of 3D RHD
atmosphere models have been constructed, in order to study surface
convection. The \textsc{cfist}-grid employing the \textsc{co5bold}-code,
studied the impact of granulation \citep[see][]{Ludwig:2009p4627,Freytag:2012p23073,Kucinskas:2013A&A...549A..14K,Tremblay2013A&A...557A...7T}.
\citet{Trampedach:2013ApJ...769...18T} established a grid of 3D RHD
models with solar metallicity. Also, efforts are being made with the
\textsc{muram}-code \citep[see][]{Vogler:2003p11832,Beeck2013A&A...558A..48B,Beeck2013A&A...558A..49B}.
We have computed the \textsc{Stagger}-grid, a grid of 3D RHD model
atmospheres \citep[hereafter Paper I and II]{Magic:2013,Magic:2013A&A...560A...8M},
and used our 3D models for several applications \citep[hereafter Paper III and IV]{Magic2014arXiv1403.1062M,Magic2014arXiv1403.3487M}.

In the present work, we perform an extensive analysis of the stellar
granulation properties based on the \textsc{Stagger}-grid. We want
to address the following key question: How do the stellar granulation
properties change for different stellar parameters? First, we explain
the granule recognition method that we have used to detect the individual
granules from the (bolometric) intensity maps of the 3D simulations
(Sect. \ref{sub:Granule-recognition}). Then, we discuss successively
the various properties of individual granules, such as their diameter
(Sect. \ref{sub:Granule-diameter}), the fractal dimension (Sect.
\ref{sub:Fractal-dimension}), the geometry (Sects. \ref{sub:Geometrical-properties_of_granules}),
the variation of the intensity, temperature, density, velocity with
granule size (Sect. \ref{sub:Granule-properties}), and finally the
properties at the optical surface (Sect. \ref{sec:optical-surface}).

\section{The granule recognition method\label{sub:Granule-recognition}}

\begin{figure*}
\includegraphics[width=61.33mm]{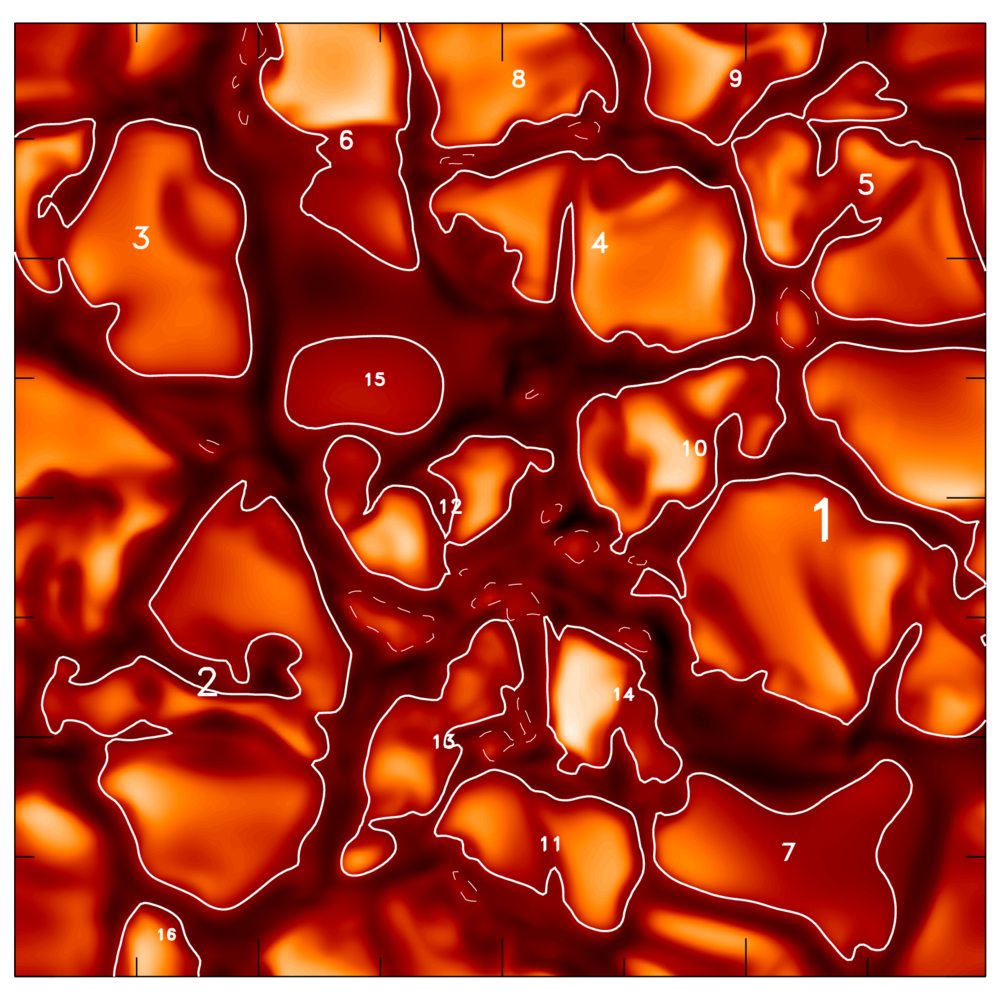}\includegraphics[width=61.33mm]{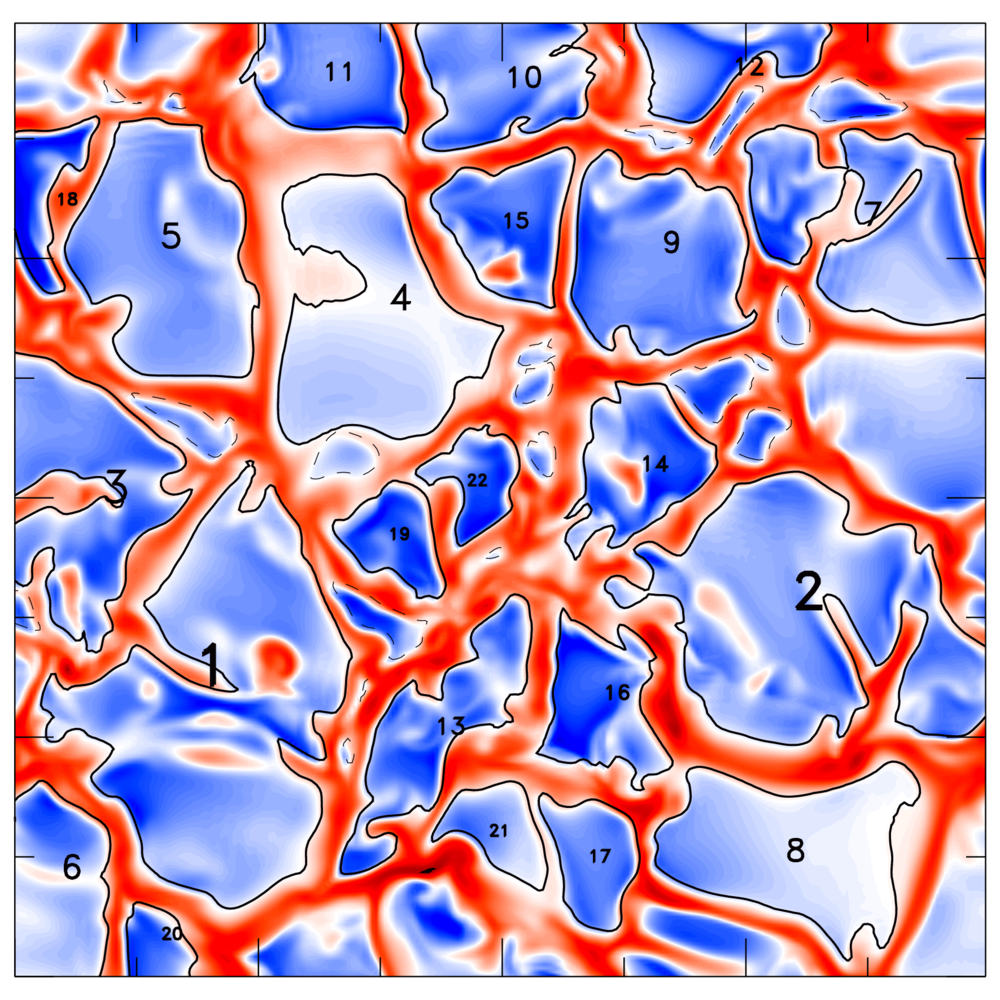}\includegraphics[width=61.33mm]{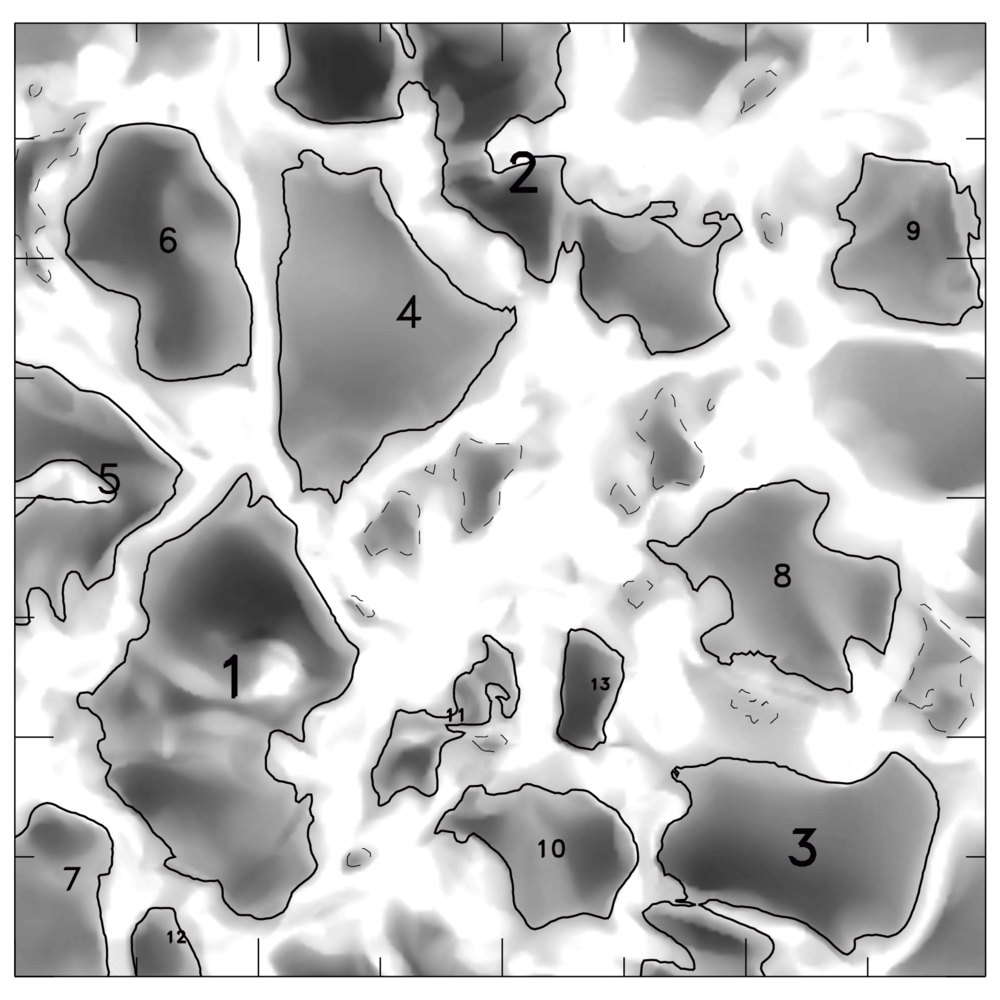}

\caption{\label{fig:solar-granules_projected}Comparison of the solar granules
detected from different variables. \emph{Left figure}: the visible
(bolometric) intensity map (orange contour); \emph{middle figure}:
the averaged vertical velocity (Eq. \ref{eq:proj_velo}; up/down:
blue/red); and \emph{right figure}: the integrated temperature excess
(Eq. \ref{eq:proj_ttc}; gray contour).}
\end{figure*}

\subsection{Detection from intensity maps\label{sub:Detection-from-intensity}}

Several methods for detecting granules in observed solar images have
been developed over the years. Classically, a single-level clip of
an intensity image is used for the granule recognition, where the
small and large features are filtered out by spatial passband Fourier
filtering. These \textquotedbl{}Fourier-based recognition\textquotedbl{}
techniques have been the most commonly applied ones in the past, and
are fast, but also inaccurate \citep[see][]{Roudier:1986p20337,Hirzberger:1997p16292}.
Another possible approach is to trace granules with a single fixed
relative intensity-level, e.g., between $0.97$ and $1.03$, as proposed
by \citet{Abramenko:2012p20140}. However, in this work we prefer
the more robust \textquotedbl{}multiple level tracking\textquotedbl{}
algorithm that was developed by \citet{Bovelet:2001p6138}. It is
a simple, yet very powerful tool to extract the granules from the
(bolometric) intensity map alone. The basic idea behind this method
is to find (granular) shapes repeatedly for decreasing intensity level
clips, thereby increasing their filling factors, until a predefined
threshold filling factor is matched. One obtains unambiguously the
granules with a single input parameter being the filling factor for
the upflows, $f_{\mathrm{up}}$. Since the latter is basically the
same for all stellar parameters with $f_{\mathrm{up}}\approx2/3$
(see Paper I), the granule-recognition is performant with the multiple
level tracking algorithm. Furthermore, we find this algorithm being
very fast, robust, and works for all stellar parameters. To demonstrate
this, we show the results for the Sun, a turnoff star, a K-giant,
and a K-dwarf as well as their metal-poor ($\feh=-3$) analogs in
Fig. \ref{fig:gran_detection}.

Following the multiple level tracking algorithm, we traced the granules
in our simulations and computed the respective filling factor, $f_{i}$,
of the considered intensity-level. We started at the relative intensity
$\bar{I}=1.2$ and decreased the intensity level down to $0.97$ in
steps of $0.01$ until the threshold with $f_{\mathrm{up}}=0.60$
was reached for all stellar parameters. We chose the threshold value
being slightly lower than the average filling factor ($f_{\mathrm{up}}\approx2/3$),
in favor of a cleaner separation between the granules. We enlarged
the intensity maps by exploiting their horizontal periodicity, and
determined the area and equivalent diameters, as well as further properties
for each granule. We dismissed very small patches and matched the
fragmented parts of granules that were located at the edge. We classified
them in two populations: small (fractional) granules and large granules
with the latter being larger than the average area. Furthermore, we
singled out dark spots and bright points located within larger granules.
Following \citet{Abramenko:2010p20160} we enhanced the contrast of
the intensity by subtracting the values $\hat{I}$ smoothed by a boxcar
average with window size of 5 pixels, and computed the root-mean-square
(RMS) for $I_{\mathrm{rms}}=I-\hat{I}$. Then, the bright points were
detected at the $2\sigma$-threshold relative to the other granules.
We performed the granule-recognition for each stored snapshot of the
time-series, thereby leading to large sample of granules for an individual
simulation (e.g., Sun $\sim3800$ for 150 snapshots every $30\,\mathrm{s}$).

\subsection{Detection from vertical velocity and temperature\label{sub:Detection-from-temperature}}

Besides the intensity map, we also detected granules from the vertical
velocity and the temperature excess with the multiple-level-tracking
algorithm. The emergent (bolometric) intensity is computed for disk-center,
therefore, it is a 2D representation of the 3D granulation structure
present in the superadiabatic region of the convection zone. To account
for the depth-dependence of the coherent granulation structures, we
averaged the vertical velocity and integrated the temperature excess
as well. 

We averaged the vertical velocity on layers of constant Rosseland
optical depth from the optical surface to the peak (maximum) of the
superadiabatic gradient (the granulation can be found in these layers
best) over ten equidistant layers in steps of $\Delta\ltaur=0.1$,
i.e. 
\begin{eqnarray}
\tilde{v}_{z}\left(x,y\right) & = & \frac{1}{10}\sum_{i=1}^{10}v_{z}\left(x,y,z_{i}\right).\label{eq:proj_velo}
\end{eqnarray}
The granules are determined on the zero velocity contour of $\tilde{v}_{z}\left(x,y\right)$.
The averaged vertical velocity is independent of any input parameter,
however, it relies on the assumption that the coherent granular structures
are given in the superadiabatic region. One could also use the vertical
velocity present at the optical surface, however, then one would neglect
the depth-dependence, while our approach by averaging over several
layers does.

We computed the temperature fluctuations for each layers by normalizing
it with the horizontal average, i.e. $\delta T_{z}=\Delta T_{z}/\left\langle T\right\rangle _{z}$.
Then, we determined the temperature excess, $\delta T=\delta T>\max(\left\langle \delta T\right\rangle _{z})/5$,
which are the (positive) temperature fluctuations above the threshold
of one-fifth of the maximum mean horizontal temperature fluctuations.
The relative threshold allows the method to work for different stellar
parameters. We integrated the temperature fluctuations from just above
the optical surface to the bottom of the simulation box with 
\begin{eqnarray}
\Theta\left(x,y\right) & = & \int_{\left\langle \tau\right\rangle =0.1}^{\mathrm{bot}}\delta T\left(z\right)dz.\label{eq:proj_ttc}
\end{eqnarray}
The granules are retrieved with the clip-level of the unity (average)
contour of $\Theta\left(x,z\right)$. In contrast to the intensity
map, this method is independent of any assumption on the filling factor.
We remark that the temperature excess, $\delta T$, is a convenient
quantity to illustrate the topology of the superadiabatic convective
cells (under-dense regions with heat-excess).

In Fig. \ref{fig:solar-granules_projected}, we compare the detected
granules from a solar simulation snapshot with the three methods.
Since the underlying variables exhibit distinct features, the recognized
granules differ slightly, in particular, the boundaries between close
granules is interpreted differently. Nonetheless, one can clearly
see that the granules correlate with bright upflowing regions of significant
$T$-excess.

The above mentioned granule detection methods assume that bright regions
in the intensity maps associate with the hotter and lighter stellar
plasma leading to the upflowing granules, which is normally fulfilled,
as we demonstrate in Sect. \ref{sub:Surface-correlations}. The remaining
dark regions in the intensity maps inherently consist of cool, dense
gas, which one usually refers as the (negatively buoyant) downdrafts.

\section{Diameter of granules \label{sub:Granule-diameter}}

\begin{figure*}
\subfloat[\label{fig:sun_geom}]{\includegraphics[width=88mm]{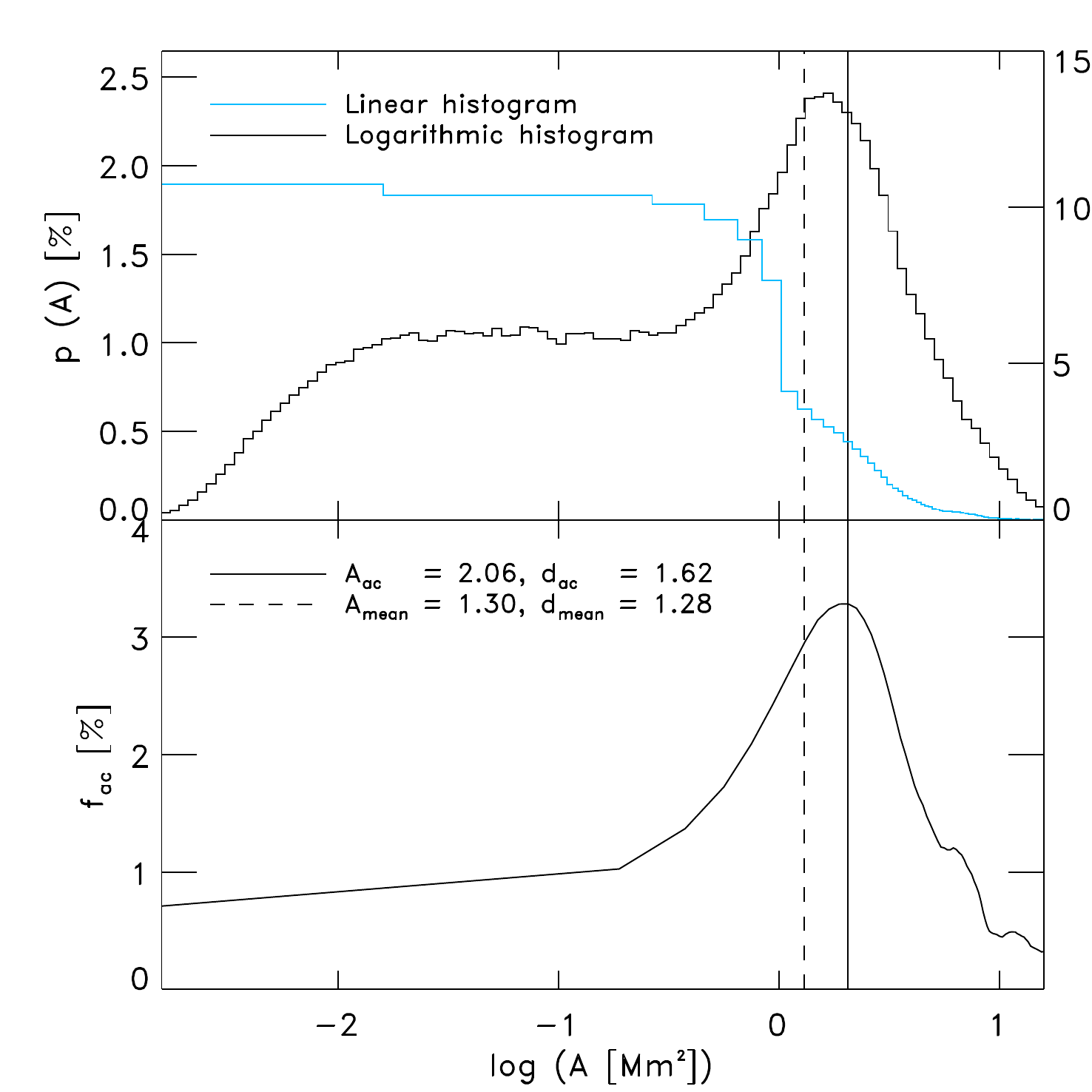}

}\subfloat[\label{fig:dgran}]{\includegraphics[width=88mm]{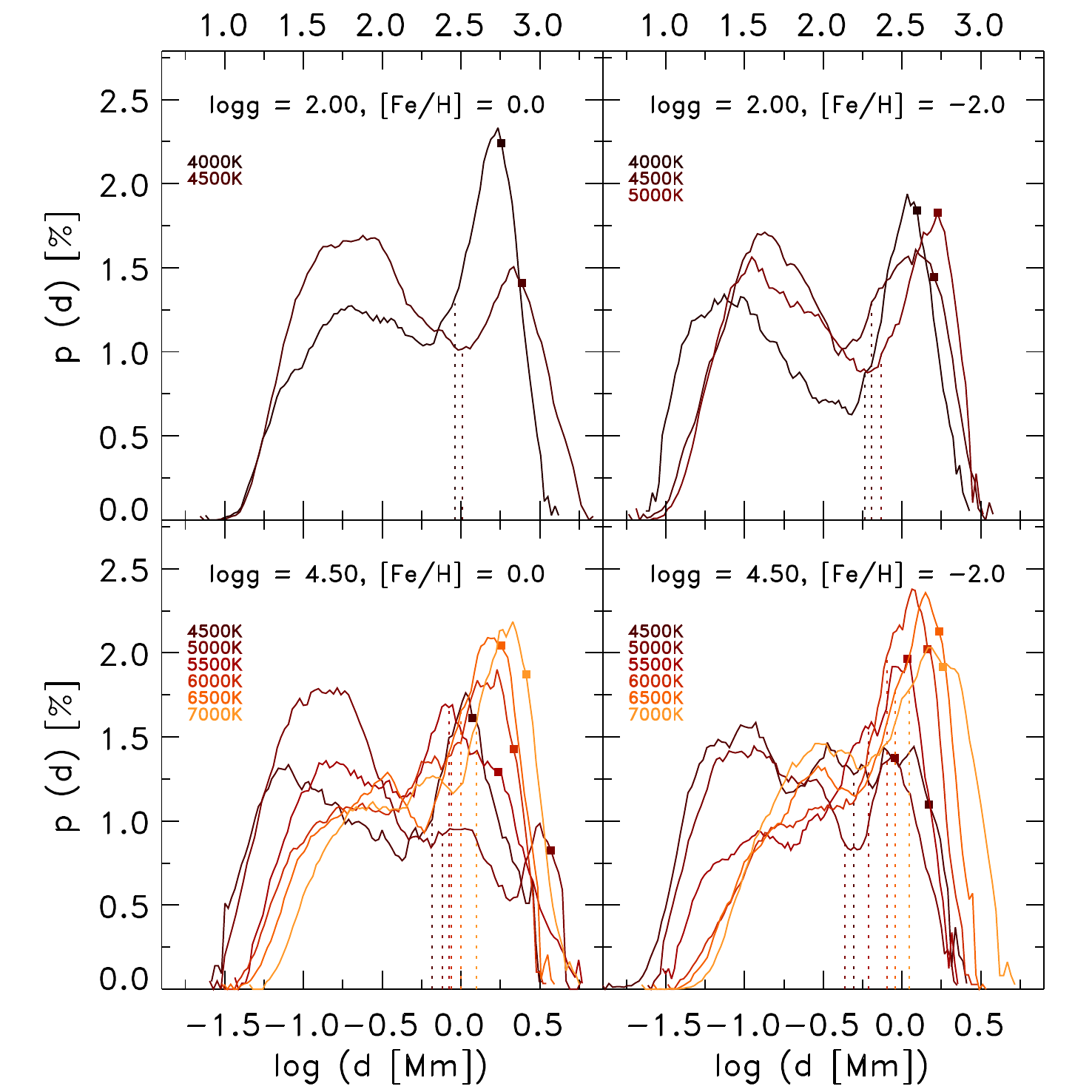}

}\caption{\emph{Left figure}: The linear and logarithmic histogram of the granule
area, $A$, (\emph{top panel}; \emph{blue} and \emph{black} line,
respectively) and the area contribution, $f_{\mathrm{ac}}$, (\emph{bottom})
derived from our solar simulation. The bin sizes for the histograms
are $0.186\,\mathrm{Mm}$ and $0.041\,\mathrm{dex}$. We indicated
the location of the mean granule area averaged over all granules (\emph{dashed})
and the maximum of $f_{\mathrm{ac}}$ (\emph{solid line}). \emph{Right
figure}: The logarithmic histograms of the granule size, $d_{\mathrm{gran}}$,
smoothed with a moving-average over 10 elements. Furthermore, we indicated
the mean granule diameters (\emph{dotted lines}, see Fig. \ref{fig:psg_dgran})
and the dominant scales, $d_{\mathrm{ac}}$ (\emph{filled square}s).
Note the difference in abscissa between the top and bottom panel.}
\end{figure*}
From the area of the granules we determined the \textit{\emph{equivalent
diameter with}} $d_{\mathrm{gran}}=2\sqrt{\agran/\pi},$ which is
the diameter of a circle that has the same area $\agran$. With the
granule size we refer to $\dgran$ or $\agran$ in the following.
Furthermore, we determined the unique \textit{\emph{barycenter}},
$\vec{x}_{\mathrm{bc}}=\sum\vec{x}_{i}A_{i}/\agran$, where the summation
runs over all pixels enclosed by the contour of the granule, and $\vec{x}_{i}$
is the vector pointing to the cell $i$, and $A_{i}$ is the pixel
area. The granule size is the first property we want to address, therefore,
we show the histogram of the granule areas\textbf{ }of the Sun in
Fig. \ref{fig:sun_geom}. The range in granule sizes is very large
(typically spanning four orders of magnitude), therefore, a histogram
considering a linear equidistant granule size for the histogram bins
would overestimate the smallest values by employing very large steps
(see Fig. \ref{fig:sun_geom}). This would result in a bottom-heavy
distribution and the misleading conclusion of a dominant a large number
of small granules \citep[see][]{Roudier:1986p20337,Hirzberger:1997p16292,Abramenko:2012p20140}.
Therefore, we advise against a linear binning of the histograms for
starkly varying quantities like the granule size, and instead consider
logarithmic granule area for the histograms. In Fig. \ref{fig:sun_geom}
the histogram exhibits a maximum close to the mean granule size, which
we refer to as the mode of granule size, i.e. $d_{\mathrm{h}}=\max\left[p\left(A\right)\right].$
The distribution around the mode of the granule size is very asymmetric
with a long tail towards smaller size. These two regimes (separated
by $d_{\mathrm{h}}$) represent on the one hand the oversized fragmenting
granules and the other hand the resulting fragments. The (fragmented)
small-scale granules were found in high-resolution solar observations
by \citet{Abramenko:2012p20140}. The fragmentation of granules is
a continuous process, therefore, the distribution of granule sizes
is also continuous, and it covers a fairly large range.

Another possibility to quantify the granule size distribution is the
\textit{\emph{area contribution function, which is given by}} $f_{\mathrm{ac}}=n_{i}A_{i}/A_{\mathrm{tot}},$
with $n_{i}$ being the number of elements within the area-bin $A_{i}$,
and $A_{\mathrm{tot}}=\sum n_{i}A_{i}$ being the total area of all
granules. The \textit{\emph{area contribution function}} is in principle
a histogram of granule size, which is weighted with the contribution
of area to the total area \citep{Roudier:1986p20337}. We noted above
that a linear granule size is overestimating the histogram for smaller
granules. The contribution function has the intrinsic advantage that
a large number of small granules contributes only little to $f_{\mathrm{ac}}$,
since their area is small. It depicts the dominant granule size, which
contributes most to the radiation (radiative losses occur mostly from
granules, which are hotter and have a larger area), independently
of the linear or logarithmic bin sizes. In Fig. \ref{fig:sun_geom}
we show the \textit{\emph{area contribution function}} resulting from
the solar simulation. We also label the \textit{\emph{dominant granule
size with the maximum, i.e.}} $d_{\mathrm{ac}}=\max\left[f_{\mathrm{ac}}\right]$,
which leads to the further definition $A_{\mathrm{ac}}=\pi\left(d_{\mathrm{ac}}/2\right)^{2}$.
Similar to the above finding with the mode $d_{\mathrm{h}}$, the
dominant granule size divides the distribution into two regimes at
a very similar value, which further supports the location of the \textquotedbl{}typical\textquotedbl{}
granule size. The decline towards larger granule sizes is similar,
but the lower part is noticeably smaller than the histogram (both
are not expected to coincide entirely due to their different definitions).
We remark that the declining tail of $f_{\mathrm{ac}}$ towards smaller
granules sizes illustrates that employing a logarithmic scale for
the histograms of the granule size is essential to yield correct conclusions
on the declining distribution of the smallest granules. We find for
the solar simulation a dominant scale of $A_{\mathrm{ac}}=2.06\,\mathrm{Mm^{2}}$,
which is $d_{\mathrm{ac}}=1.62\,\mathrm{Mm}$. Observational findings
have similar values with $d_{\mathrm{ac}}\approx1\,\mathrm{Mm}$ \citep{Roudier:1986p20337,Hirzberger:1997p16292}.
The given difference arises probably from atmospherical and instrumental
effects \citep{Stein:1998p3801}.

\begin{figure}
\includegraphics[width=88mm]{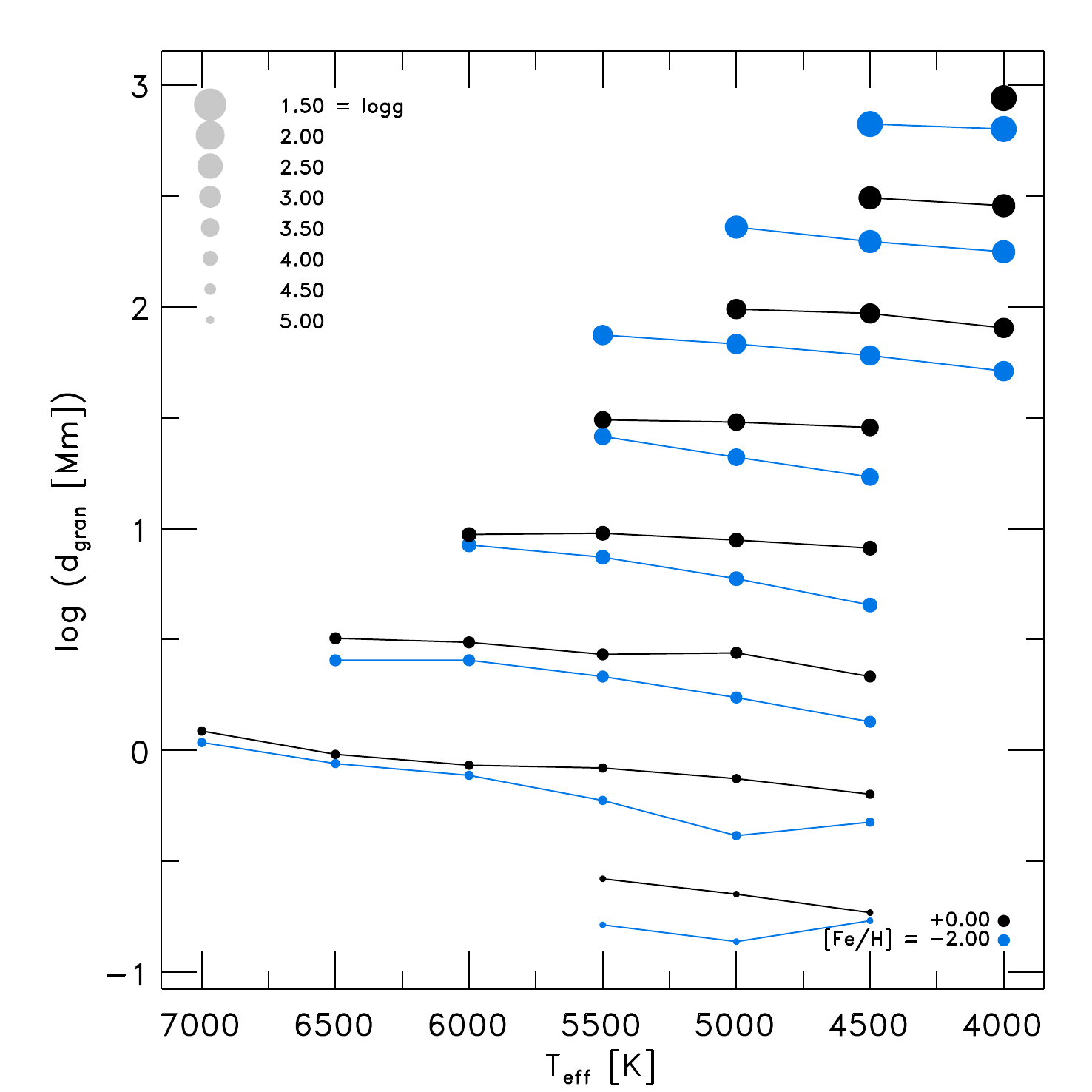}

\caption{\label{fig:psg_dgran}The mean granule size vs. effective temperature
for different stellar parameters.}
\end{figure}

In the following, we discuss the resulting granule sizes for different
stellar parameters. We show in Fig. \ref{fig:dgran} the histograms
of the granule diameters, $\dgran$. And we show the mean granule
size for different stellar parameters in Fig. \ref{fig:psg_dgran}.
For higher $\teff$ and $\feh$ the mean granule sizes are slightly
larger, while these are significantly larger for giants (lower $\logg$),
since the pressure scale height scales with the surface gravity (see
Paper I). In general, the shapes of the histogram are similar to the
solar one and also exhibit a distinct maximum in their granule sizes.
In the case of dwarfs, the peak of $d_{\mathrm{h}}$ is less pronounced
towards lower $\teff$ and we find increasingly a bimodal distribution
in the histograms of the granule diameter with a distinct second peak
from the small-scale granules, in particular, for cooler models (Fig.
\ref{fig:dgran}). The second peak at smaller granule sizes varies
with stellar parameter, so that in some cases the two peaks are reversed.
In these models the granules fragment into smaller pieces more efficiently
(see Sect. \ref{sub:Fractal-dimension}). However, most of radiation
still emerges from the larger peak, since the area contribution function
exhibits a single peak, which is located at larger granule size (see
Fig. \ref{fig:dgran}, where we have marked $d_{\mathrm{ac}}$). Furthermore,
the decline towards larger fragmenting granules is steeper with higher
$\teff$, which means that these granules are prone to disintegrate
within a smaller range of granule sizes. The lower half of the histograms
are similar despite a shift and the second peak. 

\begin{figure}
\includegraphics[width=88mm]{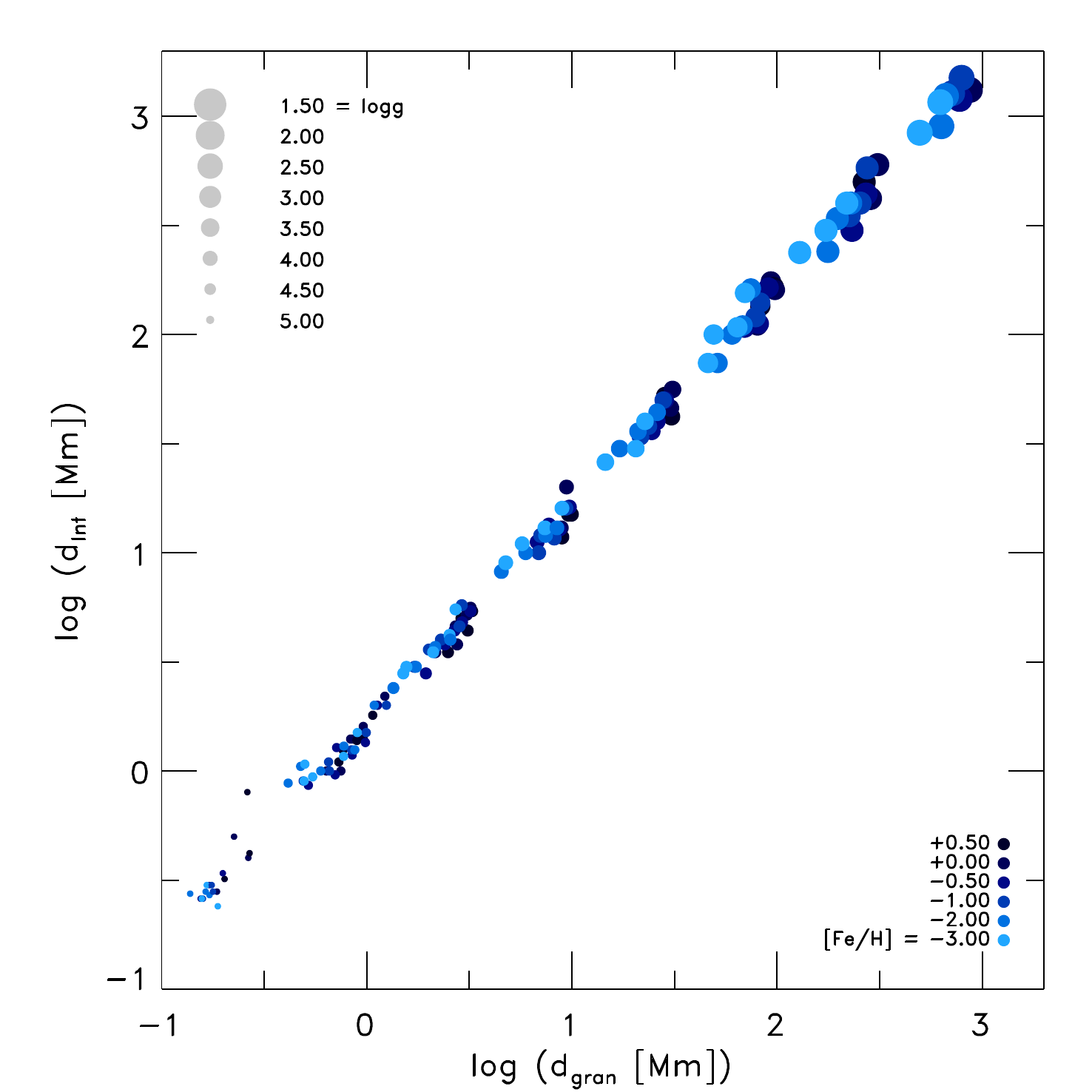}

\caption{\label{fig:dgran_vs_dint}The length-scale at the maximum of the temporally
averaged 2D spatial power-spectrum of the intensity, $d_{\mathrm{Int}}$,
vs. the mean granule size, $\dgran$, for different stellar parameters.}
\end{figure}

In Paper I, we estimated the typical granule size from the location
of the maximum of the 2D spatial power-spectrum of the intensity,
$d_{\mathrm{Int}}$. Furthermore, we found $d_{\mathrm{Int}}$ to
correlate well with pressure scale height just below the optical surface.
In Fig. \ref{fig:dgran_vs_dint}, we compare the mean granule size,
$\dgran$, with the estimated granule size from the power-spectrum
of the intensity $d_{\mathrm{Int}}$. The two correlate very well
for all of our atmosphere models.

Our findings carry some uncertainty that might be rooted in the granule
detection method or in the simulation boundaries. However, we have
confirmed that our results are robust, since these are qualitatively
similar to those by \citet{Beeck2013A&A...558A..49B}. They used also
the multiple level tracking algorithm, and found also an asymmetric
distribution exhibiting a dominant granule size with an extended tail
for the small-scale granules. Moreover, the mean granule diameters
and the filling factors they found are similar to our results.

\section{Fractal dimension\label{sub:Fractal-dimension}}

\begin{figure*}
\subfloat[\label{fig:sun_area-perim}]{\includegraphics[width=88mm]{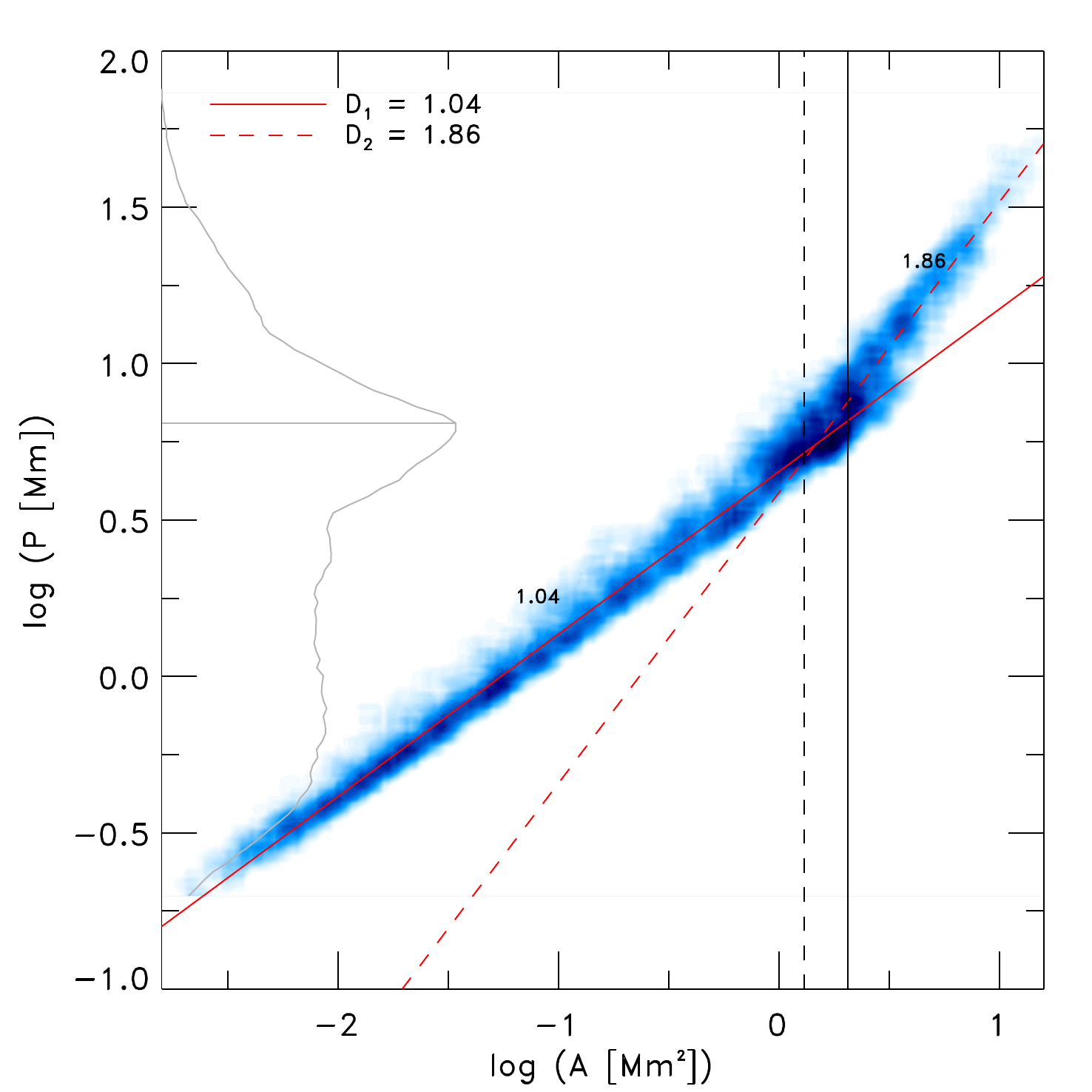}

}\subfloat[\label{fig:fracdim_d12}]{\includegraphics[width=88mm]{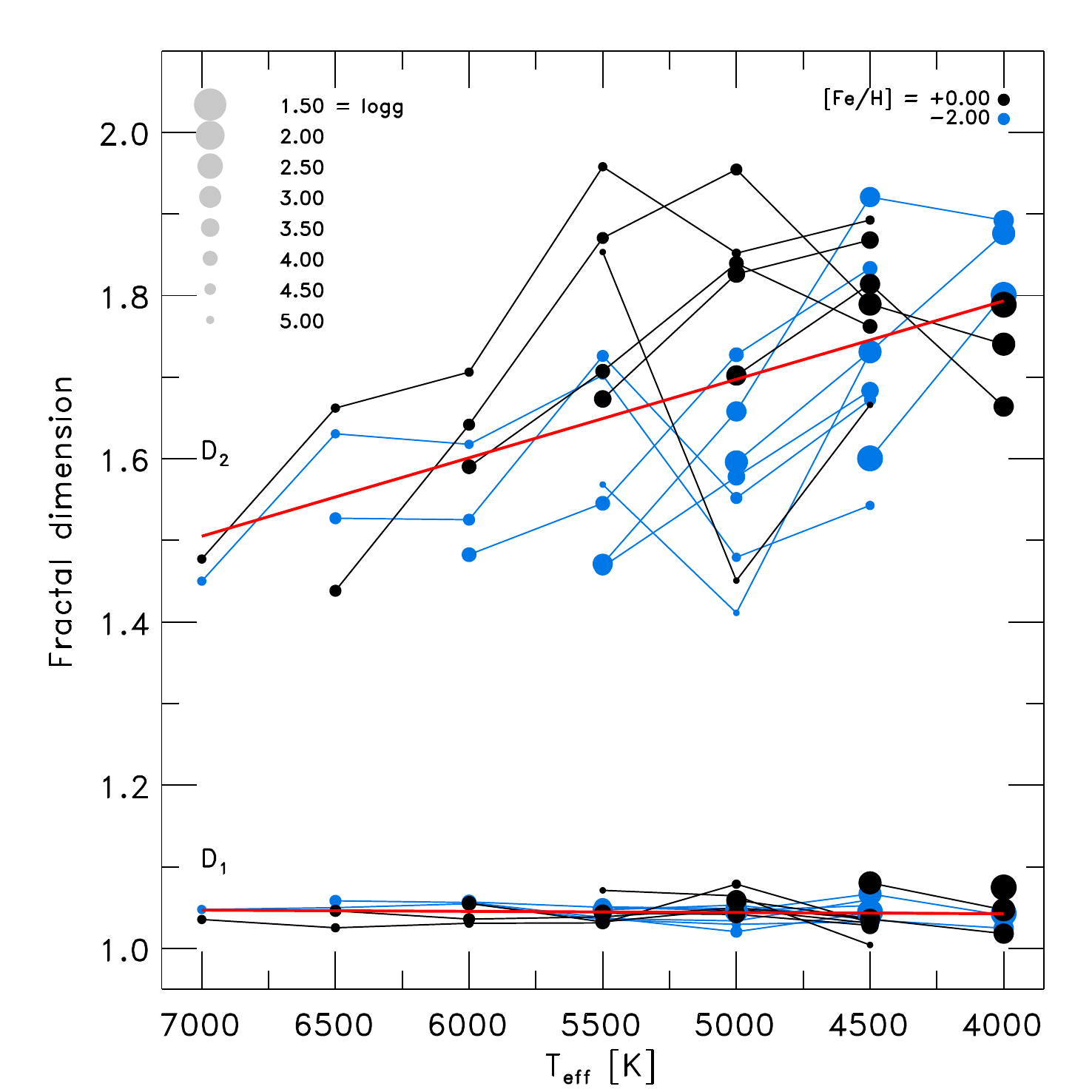}

}\caption{\emph{Left figure}: We show the (smoothed) histogram of the area-perimeter
relation determined from the solar simulation. We indicated the histogram
of the perimeter (\emph{gray line}), the location of the mean granule
area (\emph{black dashed}) and the maximum of $f_{\mathrm{ac}}$ (\emph{black
solid line}). Also the linear fit for the small-scale (\emph{red solid})
and the large granules (\emph{red dashed lines}) are also included.
\emph{Right figure}: The fractal dimensions, $D_{1}$ and $D_{2}$,
for different stellar parameter. Furthermore, we included linear fits
for $D_{1}$ and $D_{2}$, and the slopes are $\Delta_{1}\approx1.06$
and $\Delta_{2}\approx2.18$.}
\end{figure*}
The fractal dimension is a suitable, measurable value to quantify
the complexity of a geometrical shape \citep{Mandelbrot1977fgn..book.....M}.
 In the case of granules, this is given by the area-perimeter relation
\begin{equation}
P=kA^{D/2}\label{eq:area_perim_relation}
\end{equation}
with $k$ being a shape factor and $D$ the fractal dimension \citep{Roudier:1986p20337}.
In planar geometry, ideal objects have an integer fractal dimension,
e.g., circles or squares have $D=1$ (dimension of a line), but with
different shape factors $k=2\sqrt{\pi}$ and 4, respectively. However,
real objects are of fractal nature. It is an important measure for
the regularity of granules; more regular ones will have a lower $D$,
while more irregular granules will have higher area-perimeter ratios.

We show the 2D histogram of the area and perimeter determined from
the granules of the solar simulation in Fig. \ref{fig:sun_area-perim}
with a tight correlation. At the dominant granule size, we find a
distinct change in the slope of the correlation, indicating a multi-fractal
nature of granulation. Therefore, we determined two fractal dimensions
with two separate linear least-square fits. The first one is performed
for the small-scale granules ($A<\bar{A}$), and the resulting fractal
dimension is very close to unity with $D_{1}=1.04$, which means that
the smaller granules are regularly shaped. One can also consider this
in the way that for a granule that is expanding the perimeter is increasing
with the square root power of the area (same to a circle), and therefore,
these granules will be more regular shaped. The second linear fit
is performed for larger granules ($A>\bar{A}$), and the fractal dimension
is distinctively larger with $D_{2}=1.86$ (see Fig. \ref{fig:sun_geom}).
This means that large granules feature increasingly larger perimeters.

The fractal dimension has been determined from solar observations.
In agreement with our result, \citet{Roudier:1986p20337} found two
distinct fractal dimensions with $D_{1}=1.25$ and $D_{2}=2.15$ with
a Fourier-based recognition (FBR) method. Our fractal dimensions also
coincide with \citet{Hirzberger:1997p16292}, who determined for smaller
granules $ $$D_{1}\approx1.3$ and for larger granules $D_{2}\approx2.1$
(with FBR). \citet{Bovelet:2001p6138} derived $D_{1}=1.2$ and $D_{2}=1.96$.
On the other hand, \citet{Bovelet:2001p6138} determined with their
multiple layer tracking method distinctively lower values for the
fractal dimensions with $D_{1}=1.09$ and $D_{2}=1.28$. Their smaller
fractal dimension is similar to ours, however, the second one for
the larger granules is much lower. They find that the FBR method is
recognizing smaller granules as a larger single one compared to the
multiple layer tracking method. However, since we do not use a FBR
method, our results should be similar to their findings. We also performed
a single linear fit, which resulted in $D=1.10$, but the latter is
clearly insufficient to depict the larger granules (not shown). In
many of these cases, the fractal dimensions are slightly larger than
our values, which probably originates from the different granule recognition
method (FBR), but also from the reduced resolution of their observations
that include atmospheric effects.

Figure \ref{fig:fracdim_d12} shows the variation of the area-perimeter
ratio based on $D_{1}$ and $D_{2}$ for different stellar parameters.
The branching at the dominant granule size scale is always given with
a slope close to unity for the smaller granules and a steeper slope
for larger granules, in particular for lower $\teff$ and higher $\feh$.
The fractal dimension for the smaller granules is close to unity with
the average value being $D_{1}=1.05\pm0.02$, and therefore basically
universal for all simulations. This means that granules smaller than
the dominant granule size are mostly regularly shaped. Furthermore,
a value close to unity implies that the perimeter increases to the
square root with the area, i.e. $P\propto A^{1/2}$, for the smaller
granules (see Eq. \ref{eq:area_perim_relation}). The second dimensions
are clearly larger, being on average $D_{2}\approx1.7\pm0.3$ with
a significant level of scatter due to the stellar parameters, featuring
a general decreasing trend for higher $\teff$ and $\feh$, and lower
$\logg$. $D_{2}$ never exceeds 2 in every of our simulations. The
larger granules above the dominant granule size are irregularly shaped,
and $D_{2}\sim2$ translates into a linear area-perimeter relation
of $P\propto A$. This is in principle the manifestation of the fragmentation
of oversized, unstable granules. When we consider a hotter/cooler
dwarf ($\teff=7000/5500\,\mathrm{K}$, $\logg=4.5$ in Fig. \ref{fig:fracdim_d12}),
then the values for $D_{2}$ are $\sim1.5$ and $\sim1.9$. If we
compare two granules with the same (larger) area, $A_{0}$, from both
dwarfs, then the granules of the cooler dwarf will exhibit much larger
perimeters, $P_{\mathrm{hot}}(A_{0})<P_{\mathrm{cool}}(A_{0})$, i.e.
its granules will be in general more fragmented. This might be due
to the higher densities and the lower vertical velocities, thereby
shifting the balance of the characteristic length scales (see Paper
III).

The granulation pattern in our simulations exhibit a striking self-similarity
despite the large variations in the horizontal length scales and convective
flow properties (see Fig. \ref{fig:gran_detection}). This observation
is backed by the linear correlation of the area-perimeter relations,
and the similar fractal dimensions between the different stellar parameters.
Surface convection appears to operate scale-invariant over large ranges.
This is true, in particular, for the small-scale granules. Furthermore,
the branching of the two fractal dimensions is taking place at the
dominant granule size for all stellar parameters, since above the
latter the granules cannot be supported by the pressure excess and
start to fragment, thereby increasing the granule perimeter and becoming
more irregular. Therefore, the branching area between $D_{1}$ and
$D_{2}$ can be regarded as the maximal granule size, with granules
of larger sizes being unstable.

\section{Geometrical properties\label{sub:Geometrical-properties_of_granules}}

\begin{figure}
\includegraphics[width=88mm]{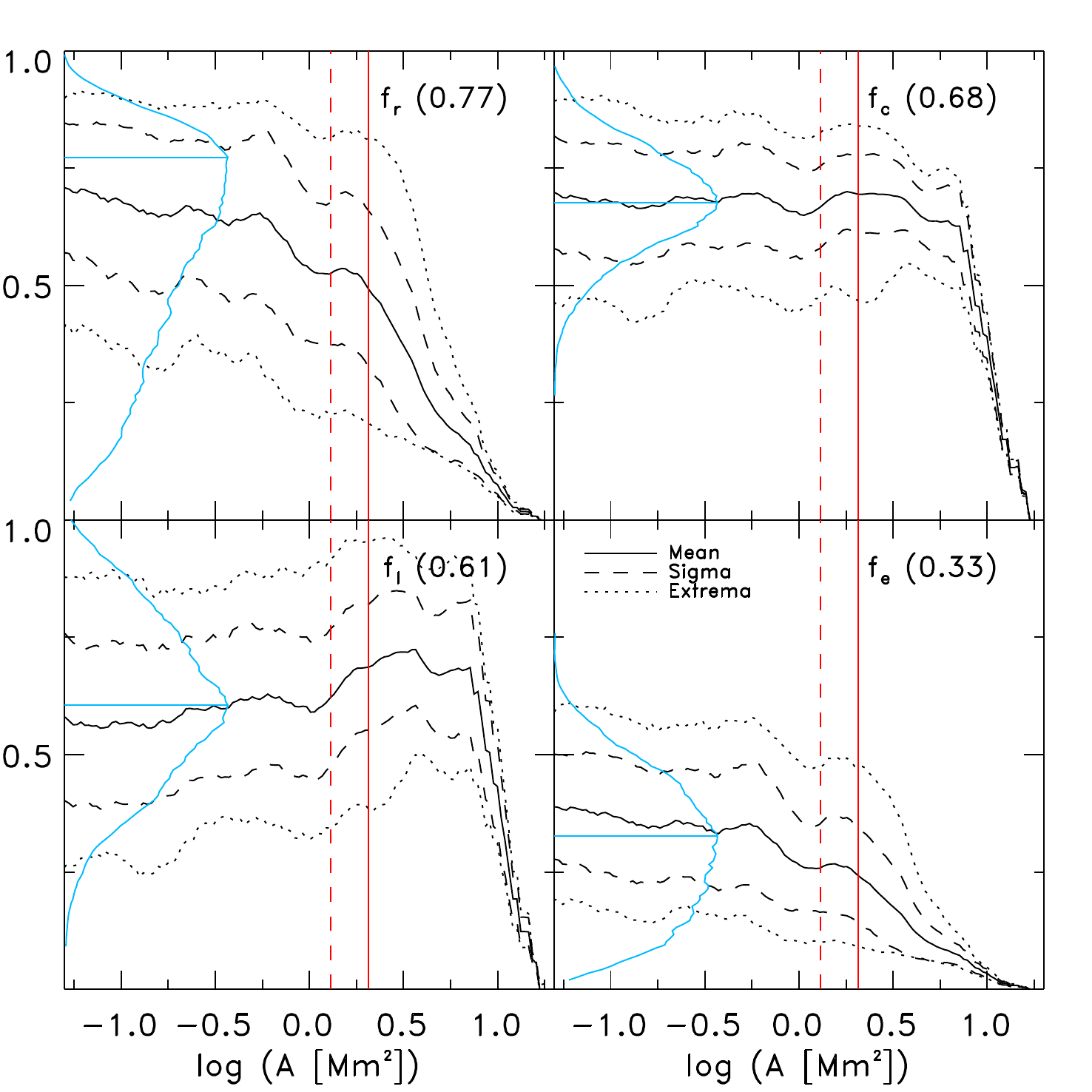}\caption{\label{fig:sun_geom_shape}The smoothed distribution of the geometrical
shape factors for roundness, circularity, elongation and ellipticity
vs. granule area derived from the solar simulation. We outlined the
mean (\emph{solid}), the standard deviation around mean (\emph{dashed})
and the extrema (\emph{dotted lines}). Furthermore, we included also
a (smoothed) histogram of the shape factor (\emph{blue lines}), in
order to render its distribution. The maximum of the latter is indicated
(\emph{horizontal blue line}). The vertical lines indicate mean and
dominant granule area (\emph{red vertical dashed }and\emph{ solid
line}).}
\end{figure}
To quantify the geometrical properties of the complex granule shapes,
we followed \citet{Hirzberger:2002p16190} and determined
\begin{eqnarray}
f_{\mathrm{r}} & = & 4\pi A/P^{2},\label{eq:isoperimeter}\\
f_{\mathrm{c}} & = & d_{\mathrm{gran}}/d_{\mathrm{MF}},\label{eq:circularity}\\
f_{\mathrm{l}} & = & w_{\mathrm{MF}}/d_{\mathrm{MF}},\label{eq:elongation}\\
f_{\mathrm{e}} & = & b/a.\label{eq:ellipticiy}
\end{eqnarray}
The \textit{\emph{roundness}} factor (Eq. \ref{eq:isoperimeter})
is the area-perimeter relation that measures the deviation from a
perfect circle, and is also known as the isoperimetric quotient. The
isoperimetric inequality, $f_{\mathrm{s}}\leq1$,  holds for any arbitrary
shape, and yields equality only for the circle. The \textit{\emph{circularity}}
factor (Eq. \ref{eq:circularity}) is the ratio between granule size,
$\dgran$, and the maximal Feret-diameter, $d_{\mathrm{MF}}$, which
is the diameter of the principal axis, i.e. the maximum diameter at
the barycenter for all degrees of rotation. It quantifies the evenness
along the boundary, where only an even shape will lead to a value
close to 1. The \textit{\emph{elongation}} factor (Eq. \ref{eq:elongation})
is determined with $w_{\mathrm{MF}}$ being the width perpendicular
to $d_{\mathrm{MF}}$, i.e. it is the aspect ratio of the principal
axis. Finally, the \textit{\emph{ellipticity}} factor (Eq. \ref{eq:ellipticiy})
is obtained by $a=\xi+(\xi^{2}-A/\pi)^{1/2}$ and $b=A/\left(\pi a\right)$
with $\xi=[\left(A/\pi\right)^{1/2}+P/\pi]/3$, and compares the shape
with an ellipse \citep[see][]{Hirzberger:2002p16190}.

The geometrical properties are shown for the Sun in Fig. \ref{fig:sun_geom_shape}.
For granules smaller than the dominant granule size, the geometrical
properties are in general very similar. Above $d_{\mathrm{ac}}$ one
finds a transition, in particular, $f_{\mathrm{r}}$ and $f_{\mathrm{e}}$
are dropping towards zero above the dominant granule size, since the
granules start to fragment and split, and the perimeter is increasing
much faster than the area ($f_{\mathrm{r}}\propto P^{-2}$). This
is in agreement with the second, larger fractal dimension, $D_{2}$,
discussed in Sect. \ref{sub:Fractal-dimension}. The shape factors
$f_{\mathrm{c}}$ and $f_{\mathrm{l}}$ are independent of the perimeter.
The histograms of the shape factors are symmetrically distributed
around a well-defined maximum with different widths. However, the
roundness factor is an exception, it exhibits a skewed distribution
that covers almost the whole range between zero and unity, and the
maximum is located at $f_{\mathrm{r}}=0.77$. As given in Fig. \ref{fig:sun_geom_shape},
the contributions arise from different granules sizes. Smaller granules
tend be in general more regular at their boundaries (larger $f_{\mathrm{r}}$)
with smooth edges, while the fragmenting granules that are larger
than $d_{\mathrm{ac}}$ have increasingly irregular, complex boundaries
(small $f_{\mathrm{r}}$) that are fringed and convoluted (see Fig.
\ref{fig:gran_detection}). The granule shapes are in overall regular,
circular shapes ($f_{\mathrm{c}}=0.68$ and $f_{\mathrm{l}}=0.61$)
independently from $\dgran$. Furthermore, the granules are quite
elongated with $f_{\mathrm{e}}=0.33$. When we compare our four shape
factors with those by \citet{Hirzberger:2002p16190}, then these are
qualitatively very similar, only the maximum of the roundness factor
is at much lower values with $f_{\mathrm{r}}=0.1$, which might due
to differences in the recognition methods. Therefore, we remark that
our solar simulation harbors a realistic granulation pattern. Since
the shape factors are very similar for different stellar parameters,
we restrict ourself to the discussion of the solar values only.

\section{Properties with granule size\label{sub:Granule-properties}}

\subsection{Intensity distribution of granules\label{sub:Intensity-distribution}}

\begin{figure*}
\subfloat[\label{fig:int}]{\includegraphics[width=88mm]{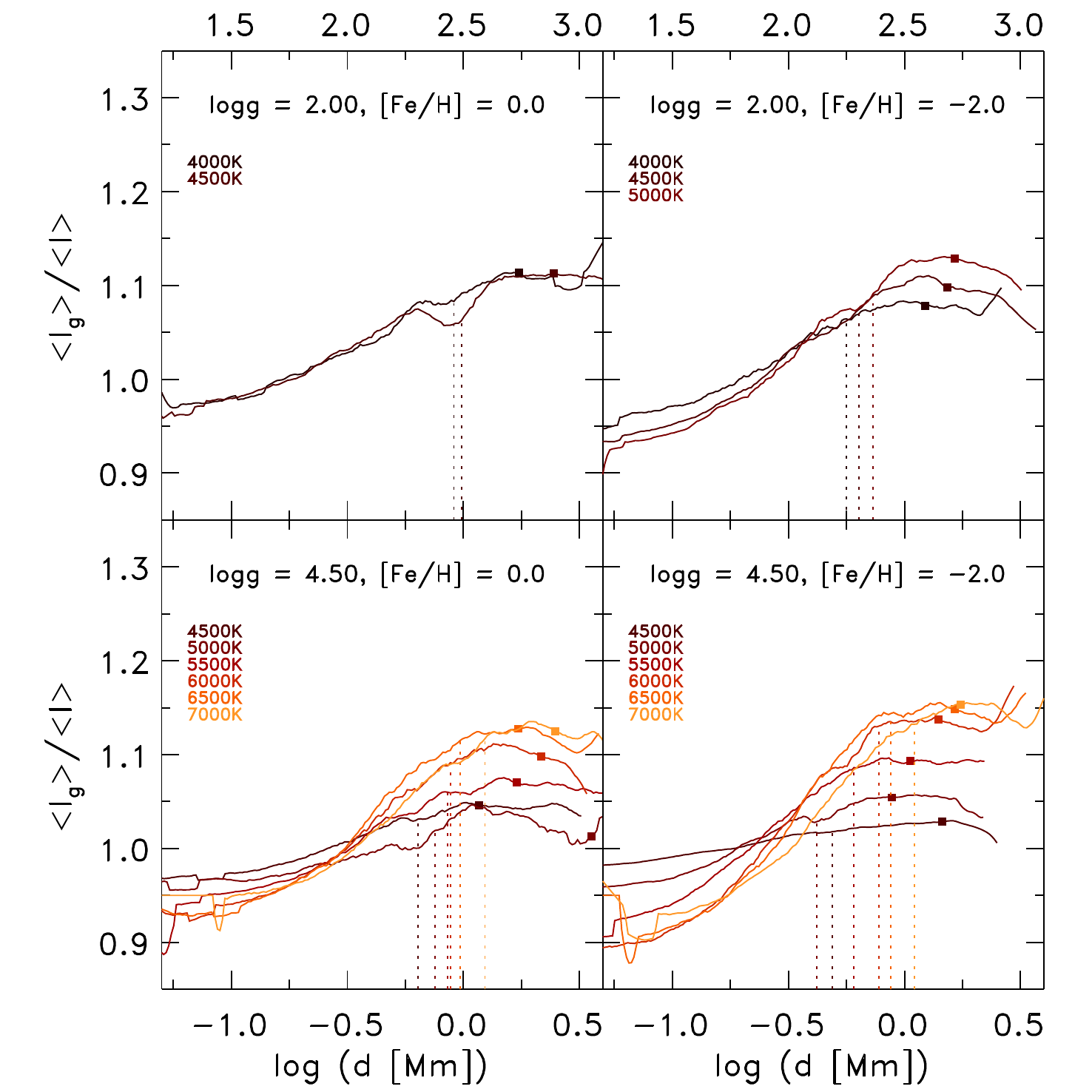}

}\subfloat[\label{fig:int_contrast}]{\includegraphics[width=88mm]{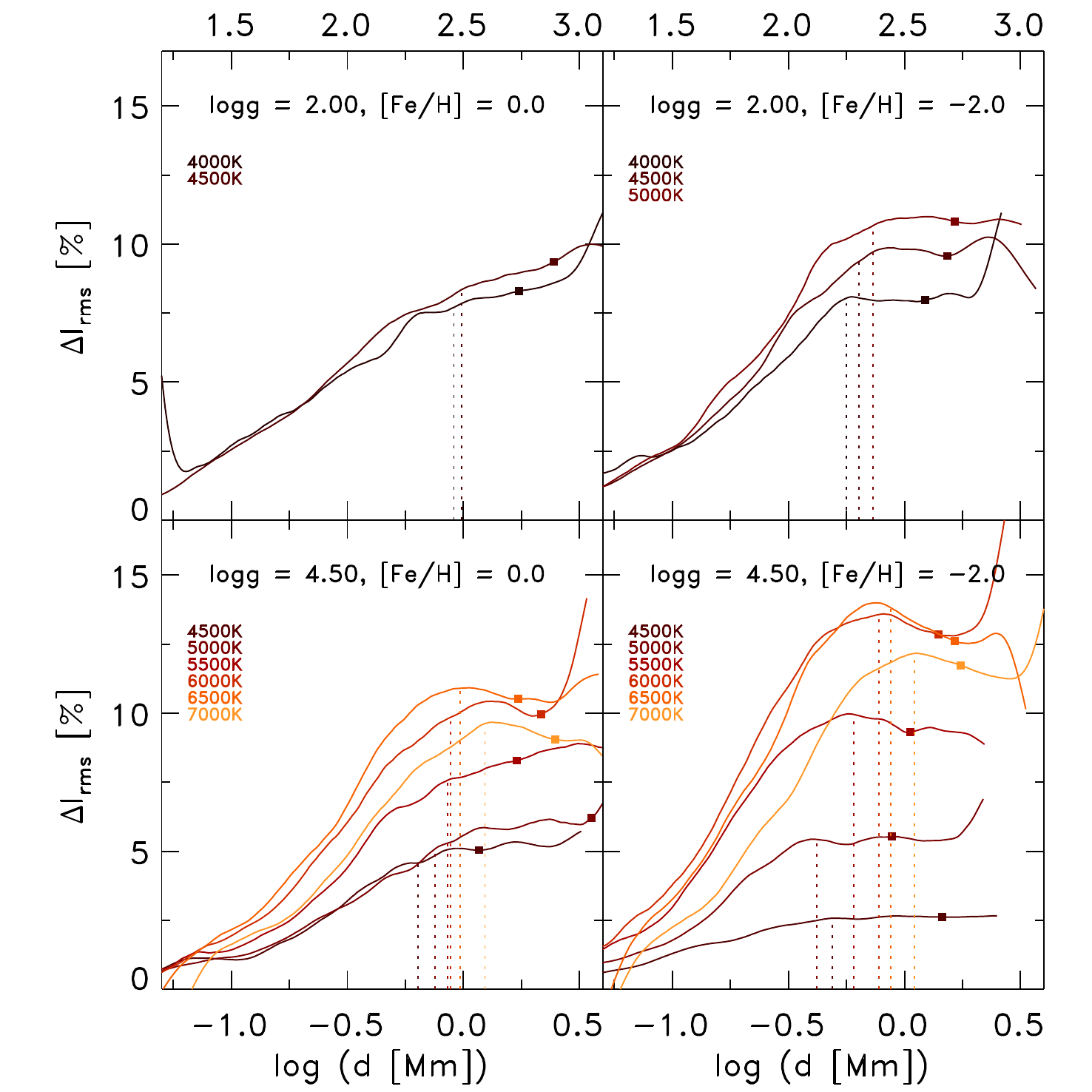}

}\caption{\emph{Left figure}: Mean normalized intensity vs. granule size. Furthermore,
we indicated the mean granule diameters (\emph{dotted lines}) and
the dominant scales, $d_{\mathrm{ac}}$ (\emph{filled square}s). Note
the difference in abscissa between the top and bottom panel. \emph{Right
figure}: Mean intensity contrast vs. granule size. Furthermore, we
indicated the dominant scales, $d_{\mathrm{ac}}$ (\emph{filled square}s).
Note the difference in abscissa between the top and bottom panel.}
\end{figure*}
In Fig. \ref{fig:int}, we show the mean (bolometric) intensities
of granules (seen at the disk-center), $\left<I_{g}\right>,$ that
are normalized by the temporal average of the entire simulation, $\left\langle I\right\rangle $,
against their granule sizes. Smaller granules are darker ($5-9\,\%$)
and larger granules are brighter. Around the dominant granule size,
$d_{\mathrm{ac}}$, one finds the brightest granules ($5-15\,\%$).
This means that the most abundant granules with sizes similar to $d_{\mathrm{ac}}$
cover most of the stellar surface and are the brightest, i.e. these
dominate the bolometric intensity not only due to their size and abundance,
but also brightness. Therefore, most of the radiative energy is lost
in these granules. The mean intensities of granules larger than $d_{\mathrm{ac}}$
are lower than the maximal, since these large fragmenting (exploding)
granules develop dark spots due to pressure excess and mass flux reversal
\citep[see][]{Stein:1998p3801}, which is then reducing the mean intensity.
\citet{Hirzberger:1997p16292} finds also a similar granule size dependence
for the mean intensity in the observed solar granules.

For higher $\teff$, smaller granules are dimmer and the larger ones
are brighter, while for different $\logg$ the changes are only subtle.
In the case of metal-poor simulations, the same small-scale granules
are darker for hotter $\teff$ and brighter for cooler $\teff$ compared
to the solar case, which correlates with the enhancement of the intensity
contrast at lower metallicity (see Paper I). Due to the lack of metals
at lower metallicity, the importance of neutral hydrogen as primary
electron-donors increases for higher $\teff$, and since the electron
density is controlling the formation of negative hydrogen -- the dominant
opacity source -- the opacity is therefore more sensitive to an increase
in temperature.

The intensity contrast of the granules vs. their size is shown in
Fig. \ref{fig:int_contrast}. The trends are similar to the intensity
with stellar parameters. The intensity contrast is lower for small
granules, typically reach a maximum at the mean granule size, and
decreasing above it. Higher intensity contrast correlates with more
complex substructures in the granules with dark spots and bright edges.
These arise due to differences in the temperature excess of the granules
originating from the granular dynamics \citep[e.g.,][]{Hirzberger:1997p16292,Stein:1998p3801}.

\subsection{Temperature and density of granules\label{sub:Temperature-density-granules}}

\begin{figure}
\includegraphics[width=88mm]{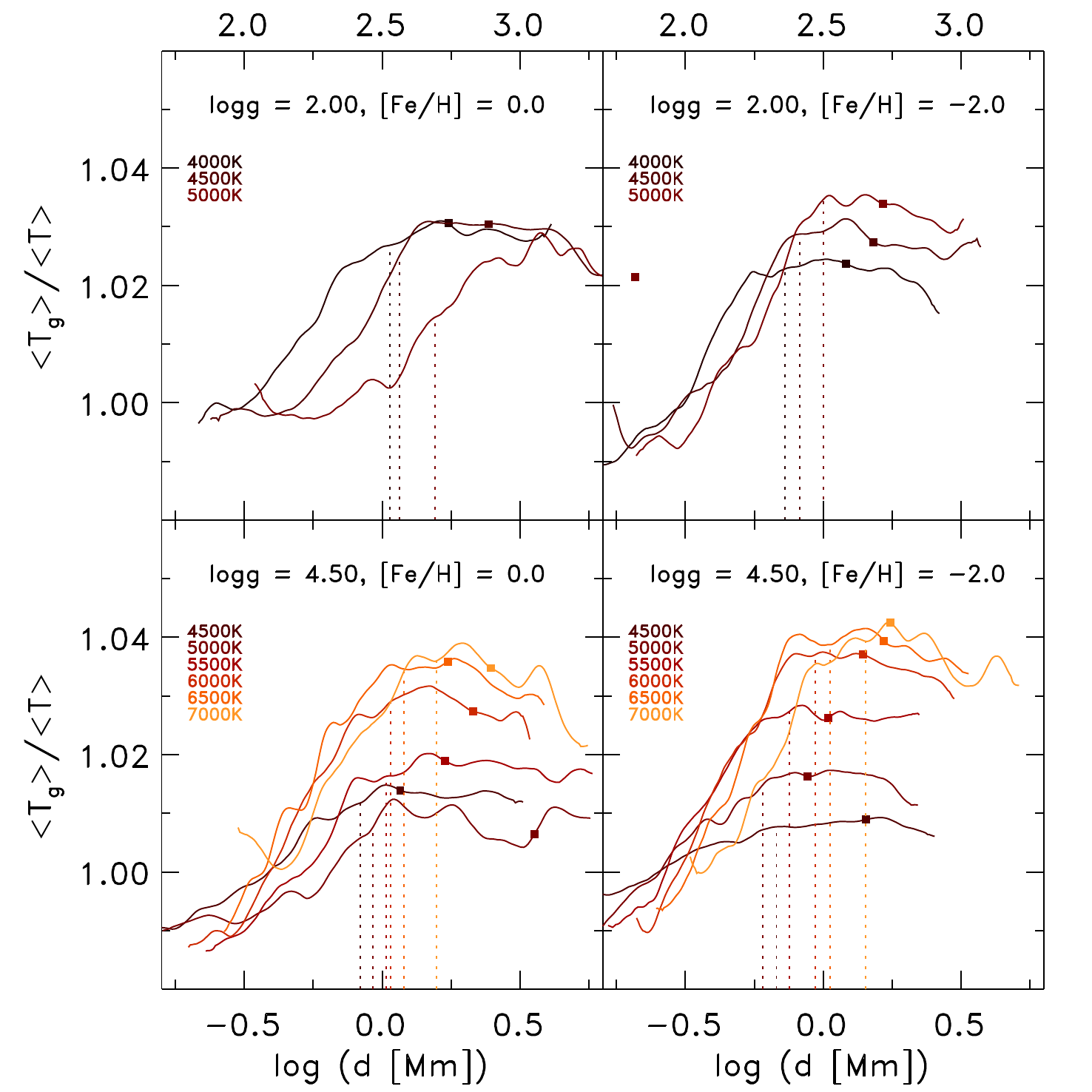}

\caption{\label{fig:gran_surf_tt}Mean normalized granule temperature vs. granule
size, which is obtained on layers of constant Rosseland optical depth.
Furthermore, we indicated the mean granule diameters (\emph{dotted
lines}) and the dominant scales, $d_{\mathrm{ac}}$ (\emph{filled
square}s). Note the difference in abscissa between the top and bottom
panel.}
\end{figure}
We averaged the temperatures and densities of the recognized granules
(Sect. \ref{sub:Granule-recognition}) on layers of constant Rosseland
optical depth at the optical surface ($\taur=1$). In Fig. \ref{fig:gran_surf_tt}
we show the temperatures against the granule sizes for different stellar
parameters. Since the densities are essentially the same as the temperatures,
we refrain from showing them. To improve the comparison the displayed
mean values of the granules, $\left<T_{g}\right>$ and $\left<\rho_{g}\right>$,
are normalized to the temporal and horizontal averages, $\left\langle T\right\rangle $
and $\left\langle \rho\right\rangle $, at the surface ($\taur=1$)
of the whole simulation. In general, larger granules feature higher
mean temperatures and lower densities. An inverse correlation between
the temperature and density is to be expected (from ideal gas law
follows $T\sim p/\rho$). The temperature excess peaks around the
mean granule diameters ($1-4\,\%$), while these are the most under-dense
granules at the same time ($1-15\,\%$). The $T$-peak and $\rho$-minimum
are increasing for higher $\teff$ and lower $\logg$. The smallest
granules exhibit lower-than-average temperatures and higher-than-average
densities, since these are small granule fragments located in the
downdrafts. Furthermore, we find a tight correlation between the mean
granule temperature and intensity with typical values around $\sim97\,\%$
with very small variation of different stellar parameters, while the
density is anti-correlated with the intensity by values around $\sim-55\%$,
but with a large variation with stellar parameters. We note that the
normalized mean thermodynamic pressure of granules, $\left\langle p_{\mathrm{th}}\right\rangle $,
exhibits very similar dependence with the granule size as the density,
while the mean entropy resembles the temperature, but on a smaller
scale (not shown).

\subsection{Velocity of granules\label{sub:Velocity-of-granules}}

\begin{figure}
\includegraphics[width=88mm]{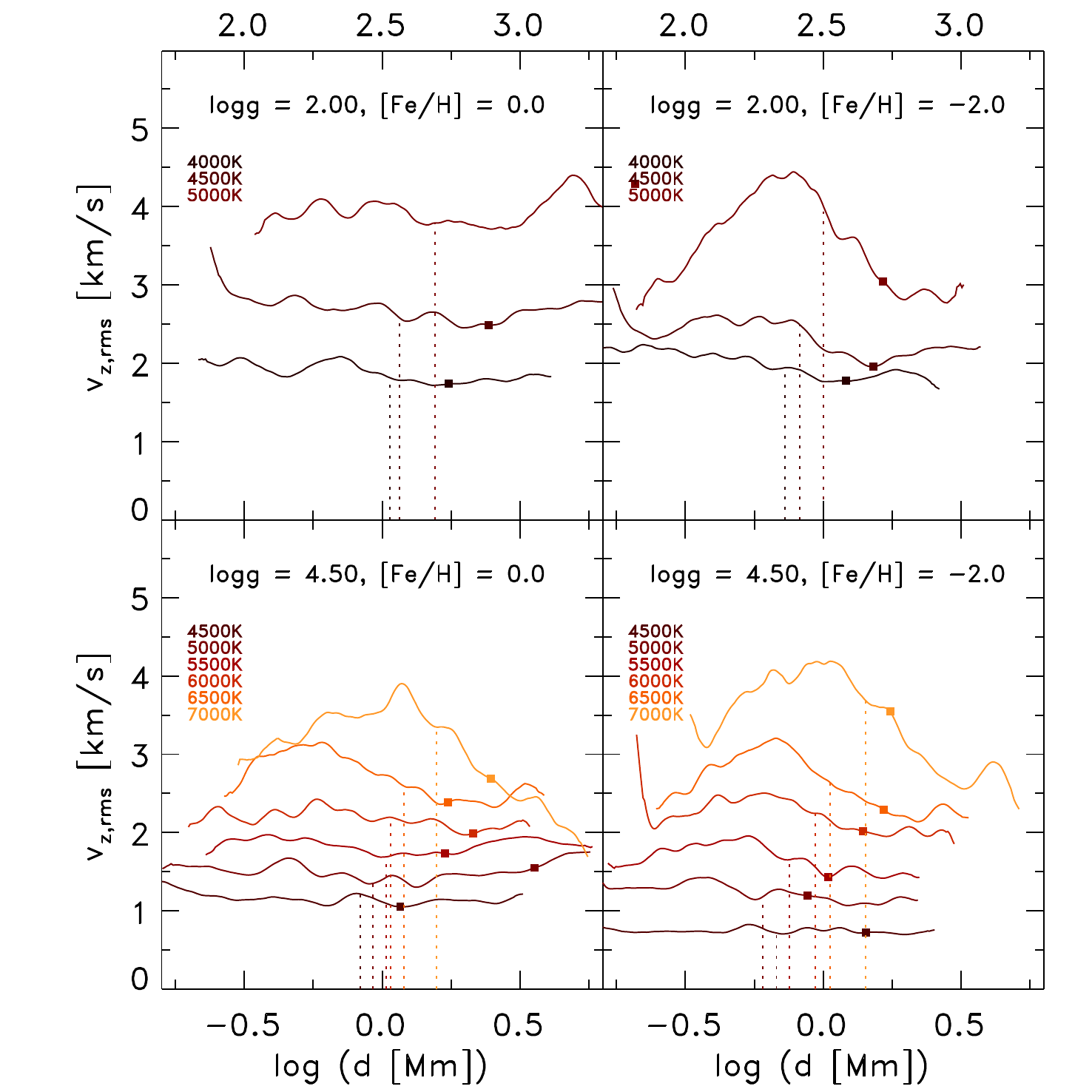}\caption{\label{fig:gran_surf_uyrms}Mean rms vertical velocity of granules
vs. granule size, which is obtained on layers of constant Rosseland
optical depth. Furthermore, we indicated the mean granule diameters
(\emph{dotted lines}) and the dominant scales, $d_{\mathrm{ac}}$
(\emph{filled square}s). Note the difference in abscissa between the
top and bottom panel.}
\end{figure}
In Fig. \ref{fig:gran_surf_uyrms}, we show the mean rms of the vertical
velocity derived for the individual granules on layers of constant
Rosseland optical depth at the optical surface. The rms velocity are
increasing for higher $\teff$, lower $\logg$ and higher $\feh$.
These are in general flat for lower $\teff$, while for higher $\teff$,
one can find a distinct peak close to the mean granule diameter. We
have seen above (Sect. \ref{sub:Temperature-density-granules}) that
these granules with mean diameters have lower densities due to higher
temperatures, therefore, these lighter granules will experience a
larger buoyancy acceleration. We remark that the characteristic variations
of the rms velocities arise mainly from the upflowing material. We
find that the lower mean densities around the mean granule diameters
are not decreasing the mean upwards directed vertical mass flux, since
the higher velocities are raising the upwards mass transport.

\section{The optical surface\label{sec:optical-surface}}

\subsection{Corrugation of the optical surface\label{sub:Corrugation-optical-surface}}

\begin{figure}
\includegraphics[width=88mm]{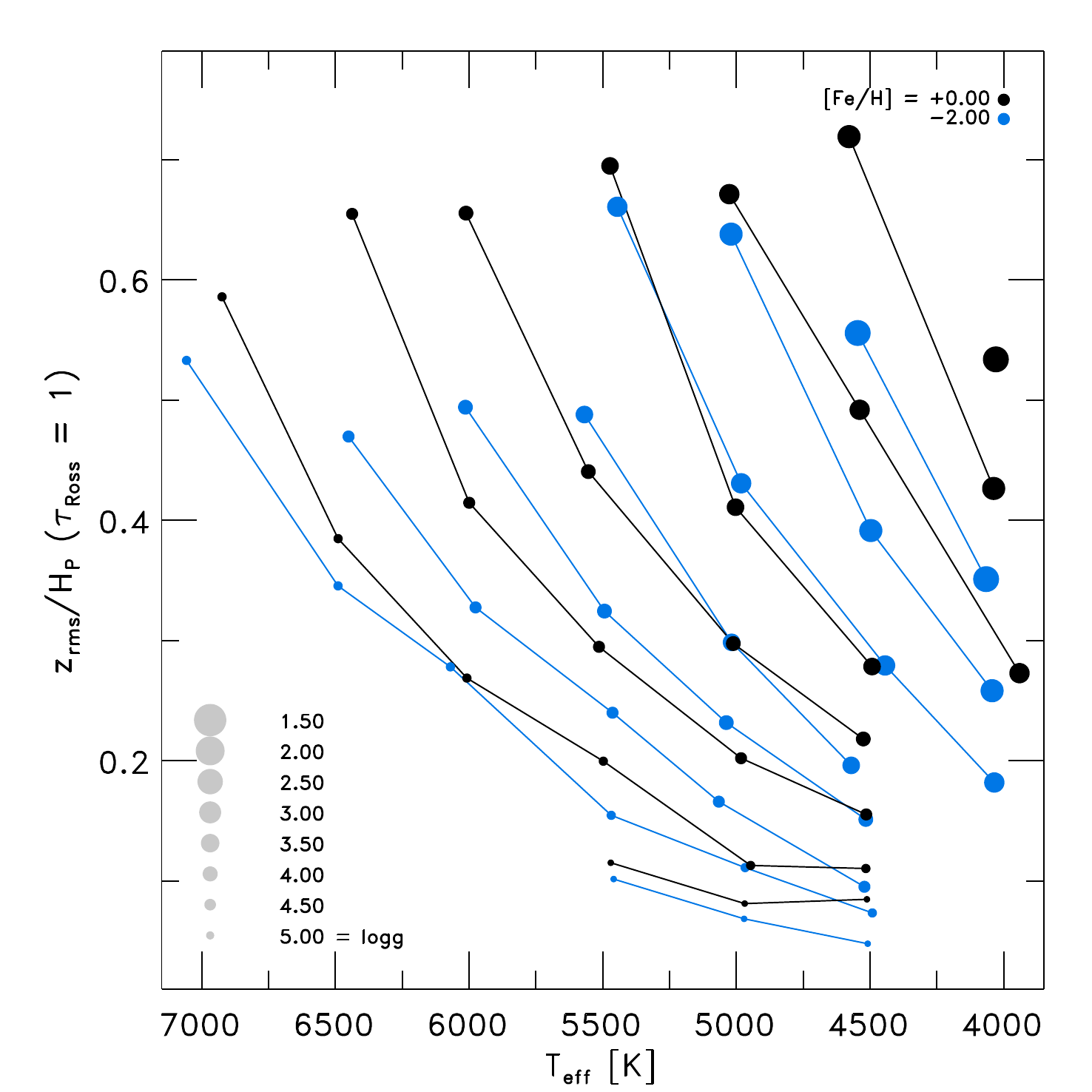}\caption{\label{fig:ov_tdepth}Overview of the rms deviation of the geometrical
depth for optical depth unity that is normalized with the pressure
scale height for different stellar parameters.}
\end{figure}
\begin{figure*}
\hspace{25mm}$\teff=5777\, K,\,\logg=4.44$\hspace{52mm}$\teff=6500\, K,\,\logg=4.00$

\includegraphics[bb=0bp 100bp 1148bp 950bp,clip,width=88mm]{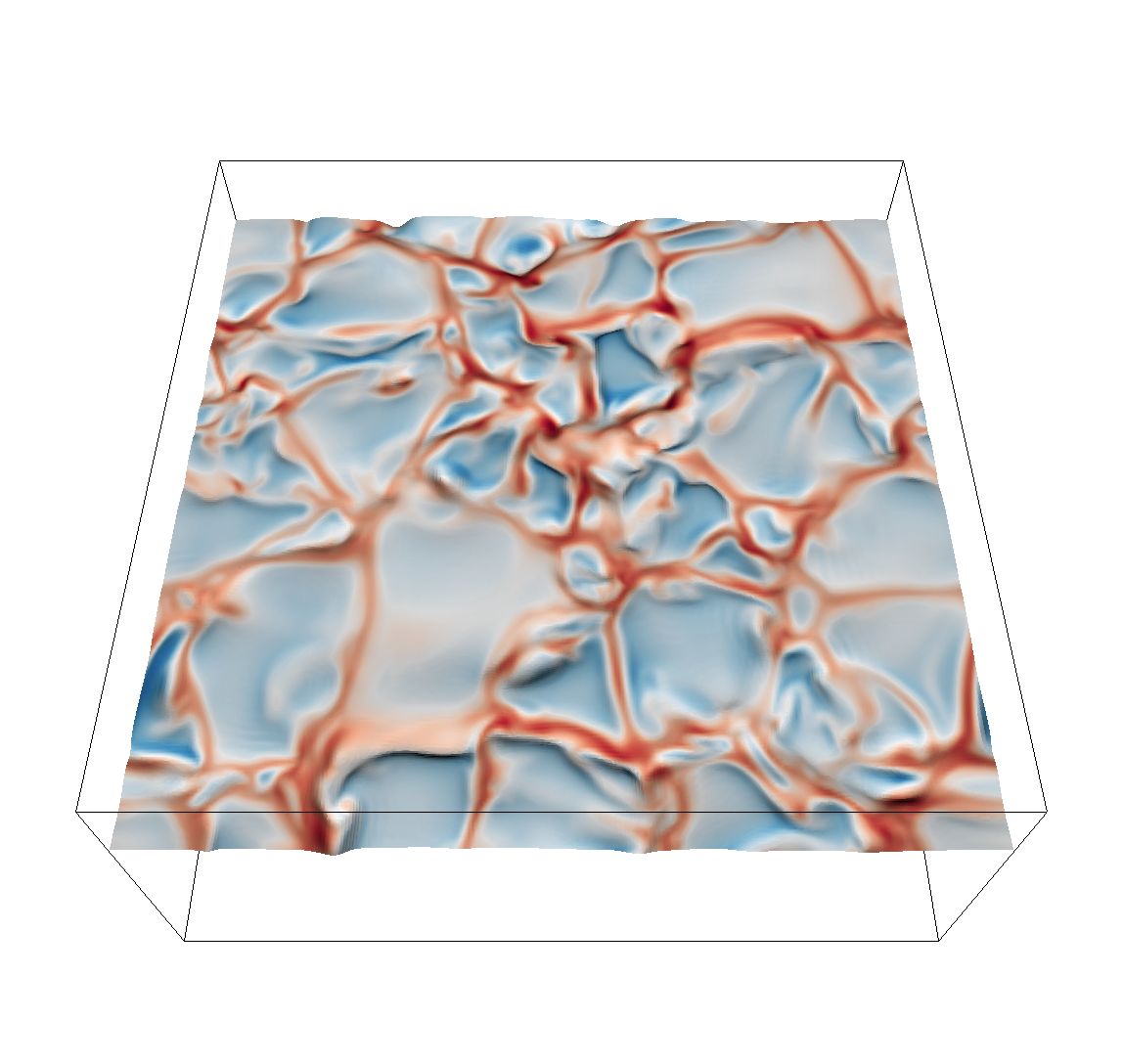}\includegraphics[bb=0bp 100bp 1148bp 950bp,clip,width=88mm]{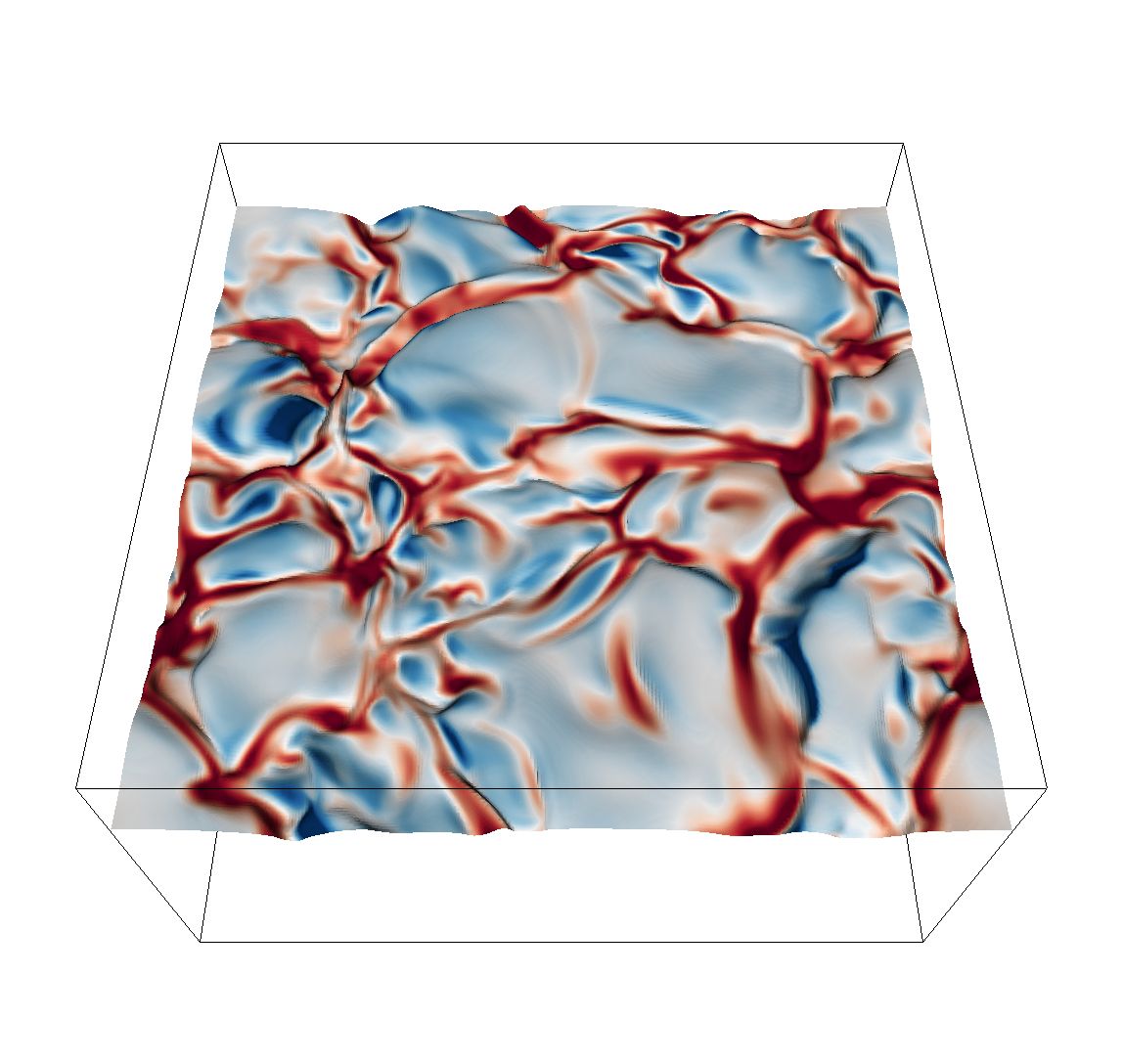}

\hspace{25mm}$\teff=4500\, K,\,\logg=2.00$\hspace{52mm}$\teff=4500\, K,\,\logg=5.00$

\includegraphics[bb=0bp 100bp 1148bp 950bp,clip,width=88mm]{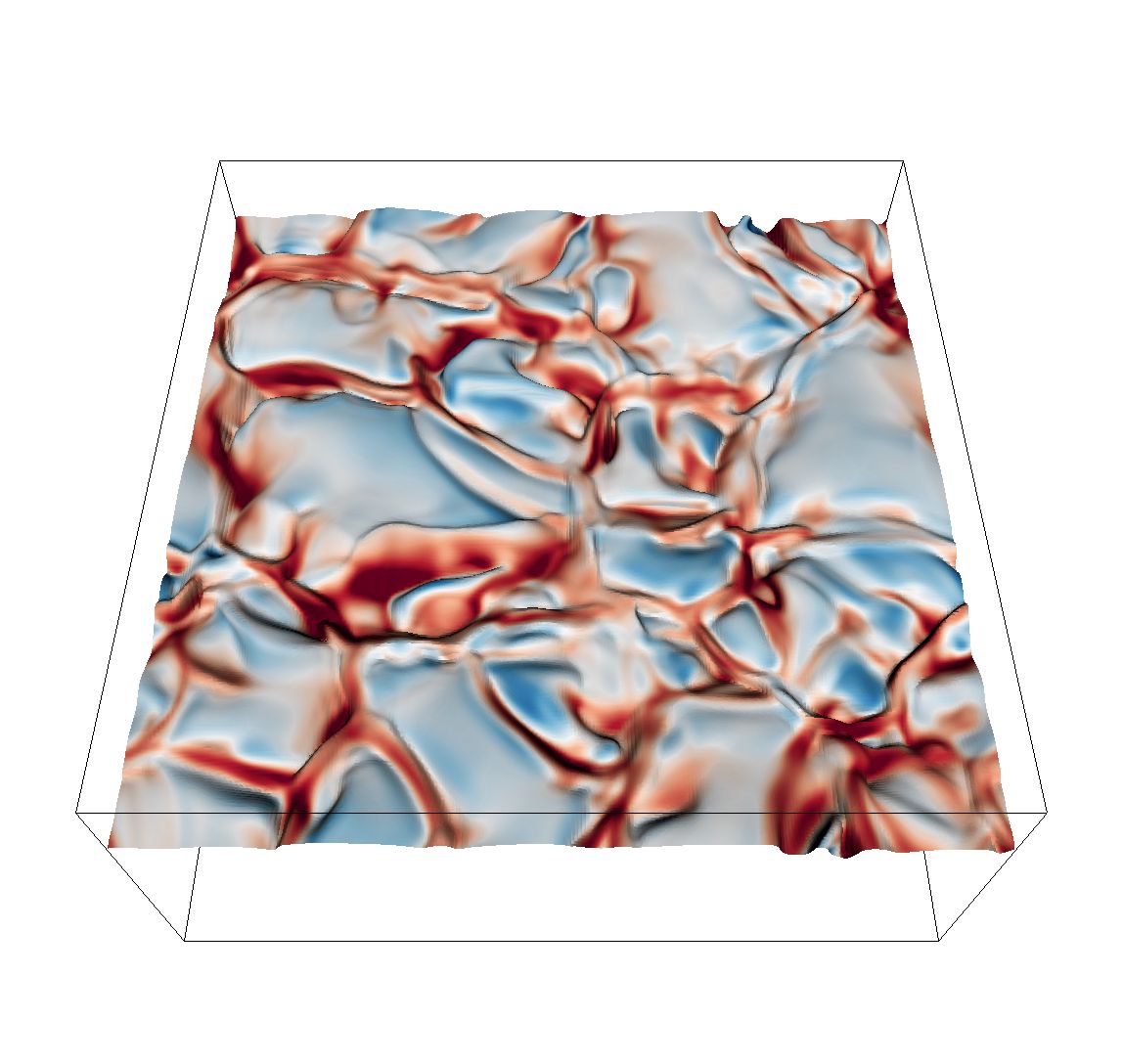}\includegraphics[bb=0bp 100bp 1148bp 950bp,clip,width=88mm]{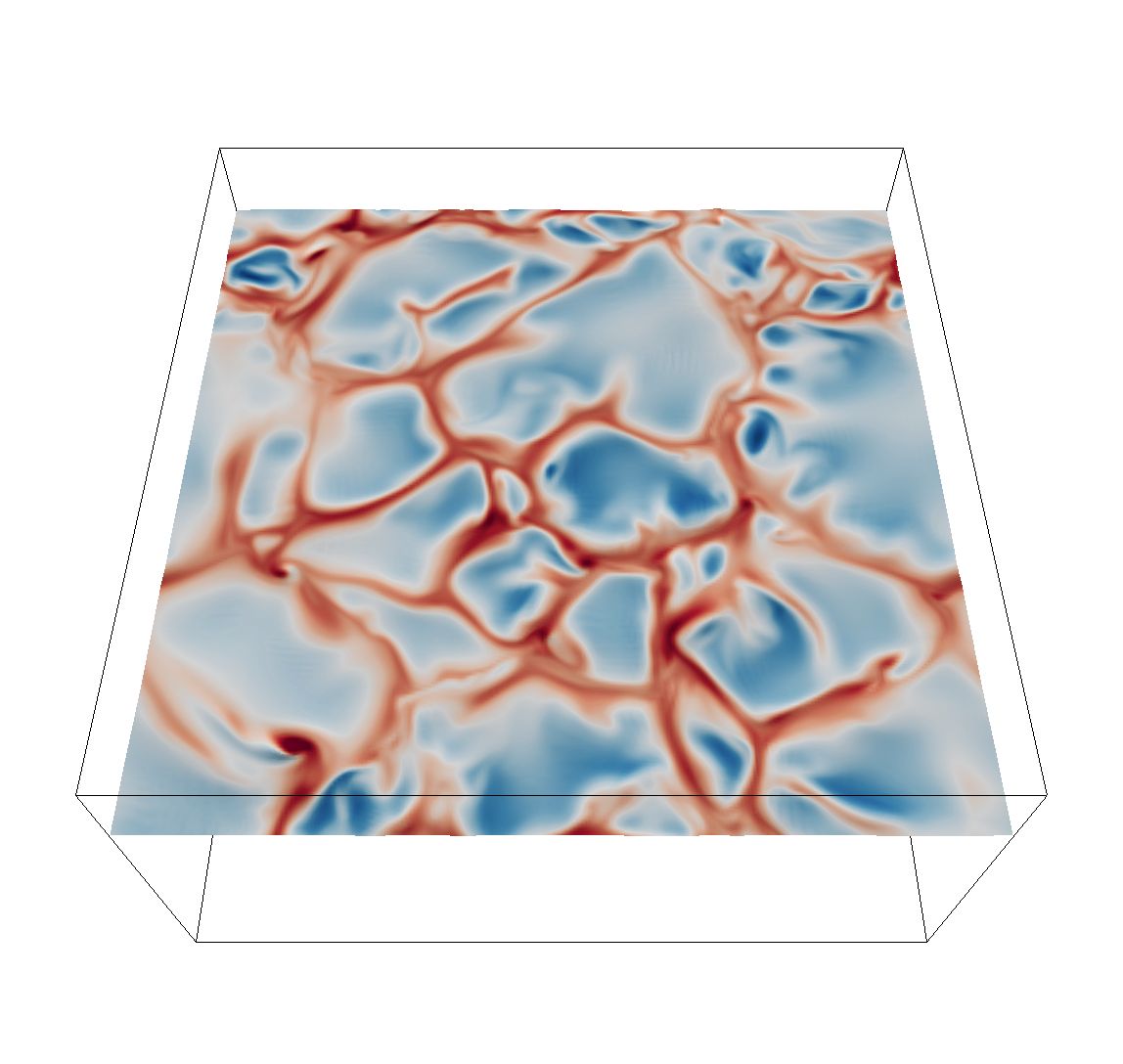}

\caption{\label{fig:optical_surfaces}The corrugated optical surface with the
viewing angle $\mu=0.6$, including the vertical velocity to illustrate
the up- and downflows (\emph{blue} and \emph{red}; each with a range
of $8\,\mathrm{km/s}$) for a selection of stars: Sun ($5777\,\mathrm{K}/4.44$),
turnoff ($6500\,\mathrm{K}/4.0$), K-giant, K-dwarf with solar metallicity.}
\end{figure*}
The optical surface is defined as the layer with optical depth unity
($\taur=1$), and marks the photospheric transition boundary to the
outside. The optical surface is corrugated depending on whether one
is considering a region above a granule or one above the intergranular
lane, since the local optical depth depends primarily on the integral
of the temperature and its gradient. Therefore, we observe the emitted
light above granules from higher layers, while the radiation from
the intergranular lanes originates from slightly deeper geometrical
layers. In Fig. \ref{fig:optical_surfaces}, we show the optical surfaces
for four different stellar parameters, encompassing the Sun, a turnoff
star, a K-giant and a K-dwarf. Furthermore, we also illustrate the
vertical velocity at the optical surfaces, showing that the downflows
are located in the intergranular lanes, while the granules are flowing
upwards at the surface. 

The level of corrugation differs for the different stellar parameters.
The level of corrugation can be quantified with the temporal averaged
rms deviation of the geometrical depth for the layers of constant
optical depth unity, i.e. $\left\langle z_{\mathrm{rms}}\left(\taur=1\right)\right\rangle $.
The solar simulation is slightly corrugated with $\sim33\,\mathrm{km}$,
which is close to the value found by \citet{Stein:1998p3801} with
$\sim30\,\mathrm{km}$. Compared to the solar radius this is a very
small relative variation: $\sim5\times10^{-3}\,\%$. In comparison
the Earth surface has a tolerance of $0.17\,\%$ from a spheroid,
which is $\sim30$ times larger than the (quiet) Sun. The turnoff
and giant simulation exhibit much larger corrugated optical surfaces
compared to the Sun with $300$ and $23\,000\,\mathrm{km}$ respectively,
while the dwarf model has a very smooth optical surface with $3\,\mathrm{km}$.
One can estimate that the turnoff star has approximately twice the
solar radius, while the K-giant is twenty times higher, making the
relative variations $0.02$ and $0.17\,\%$, respectively. The K-dwarf
would have half of the solar radius and a very small relative variation
with $\sim1\times10^{-3}\,\%$. 

To illustrate the systematic variation of the corrugation, we overview
the standard deviation of $\left\langle z\left(\taur=1\right)\right\rangle $,
which is normalized by the (total) pressure scale height in Fig. \ref{fig:ov_tdepth}.
The corrugation increases primarily with lower surface gravity and
higher effective temperatures, since from hydrostatic equilibrium
($dp/dz=\rho g$) follows $dz\propto1/g$ and the dominant negative
hydrogen opacity source is very temperature sensitive ($\kappa_{\mathrm{H}^{-}}\propto T^{10}$;
\citealt{Stein:1998p3801}). Furthermore, at higher metallicity the
corrugations are also larger due to the lower densities.

\subsection{Surface velocity correlations\label{sub:Surface-correlations}}

\begin{figure*}
\includegraphics[width=88mm]{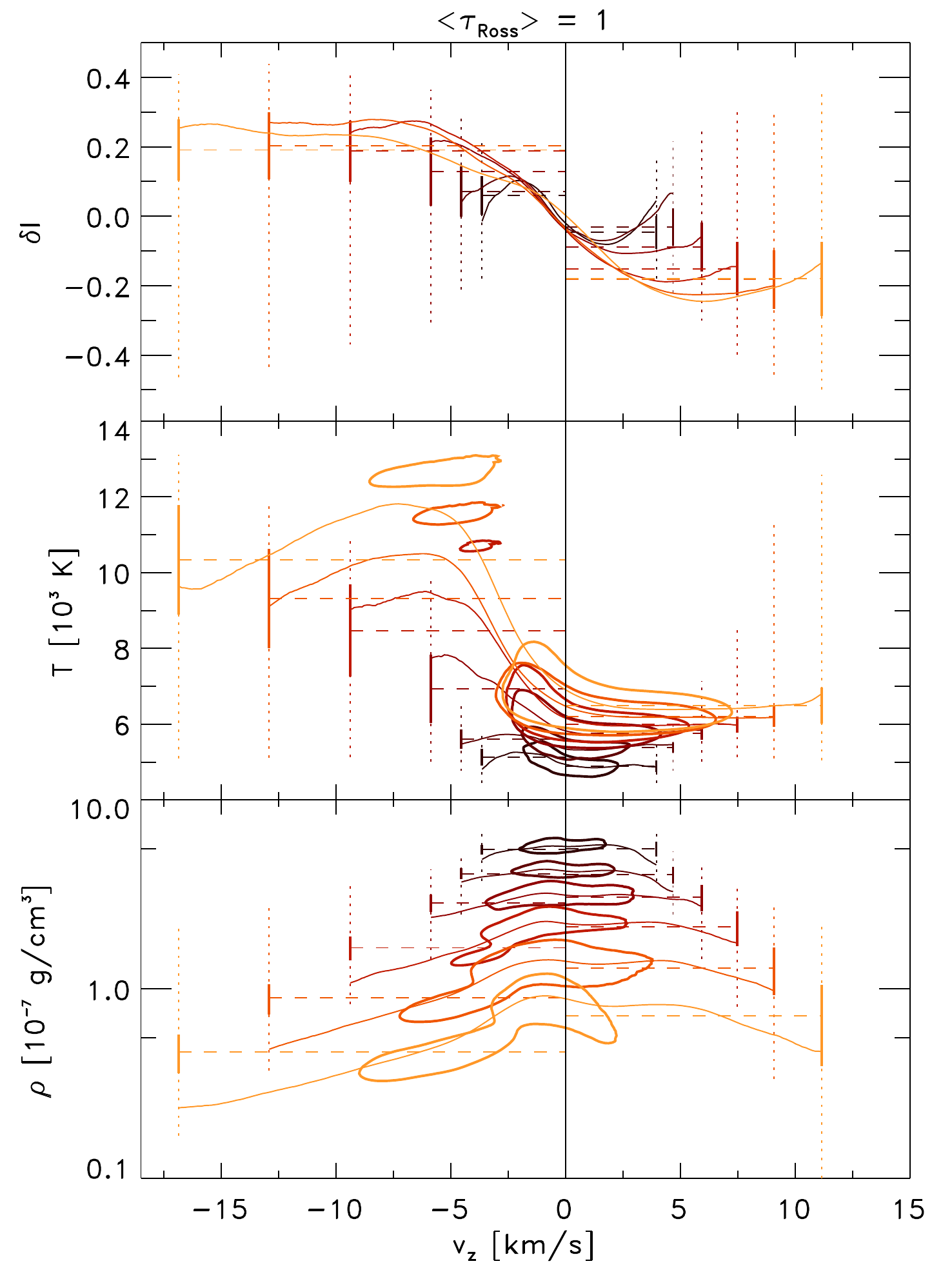}\includegraphics[width=88mm]{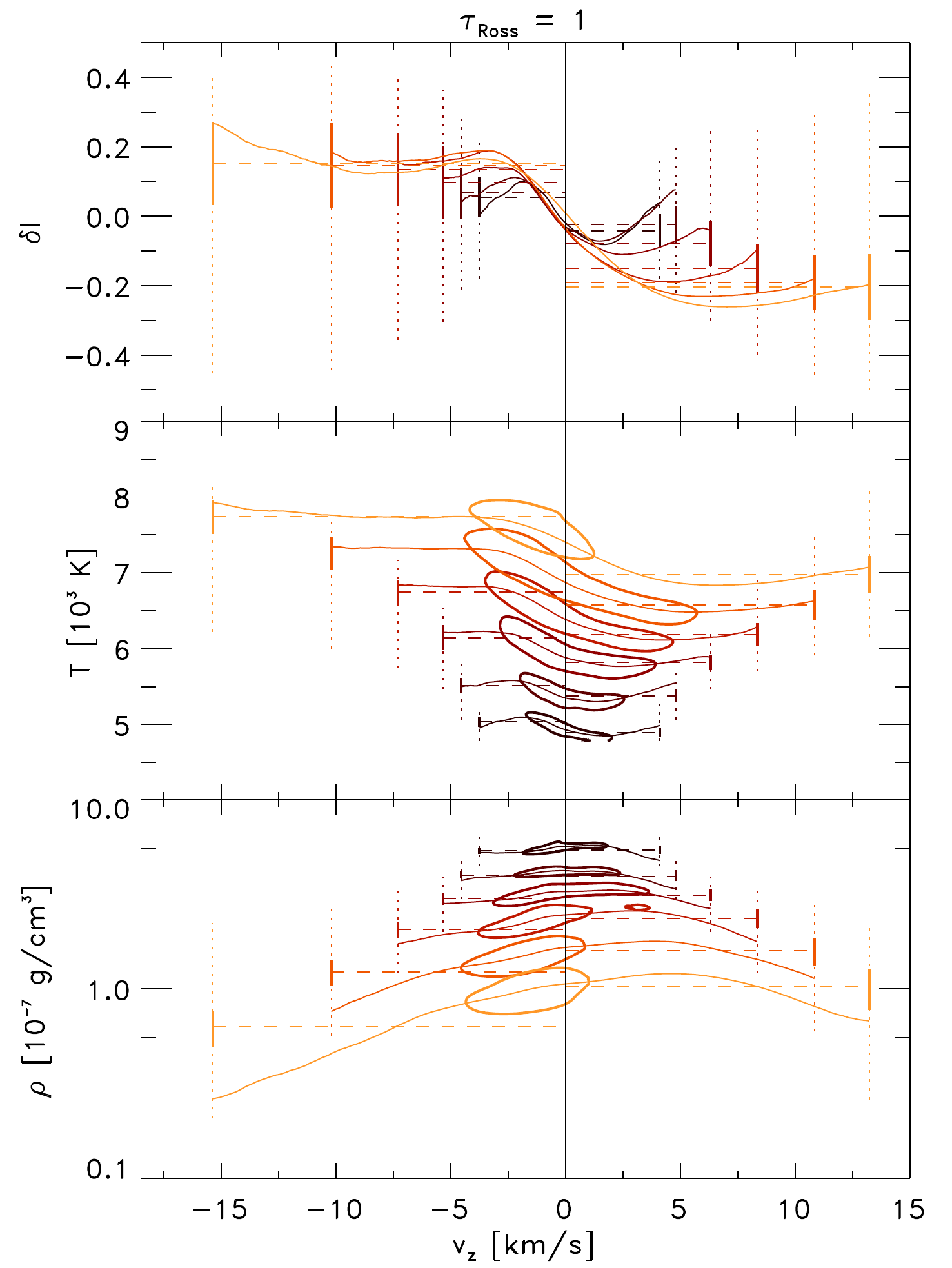}

\caption{\label{fig:surface_velocity_correlation}Correlation of the relative
intensity fluctuations, temperatures, and densities with the vertical
velocities at the optical surface for different stellar parameters
(top, middle and bottom panel respectively) shown by the distribution
of their histogram (normalized histogram at 0.2 is shown with thick
contour lines). We indicated the mean value (solid line), and for
the up- and downflow separately their mean (horizontal dashed line),
range (vertical dotted line) and standard deviation (vertical solid
line). Furthermore, we show the geometrical averages taken at the
height with $\left\langle \taur\right\rangle =1$ (left panel) and
the averages on layers of constant optical depth $\taur=1$ (right
panel).}
\end{figure*}
In order to study the surface properties more closely, we determined
the temporally averaged (2D) histograms for the temperature, $T$,
density, $\rho$, and intensity fluctuations, $\delta I$, as a function
of vertical velocity at the optical surface on layers of geometrical
depth ($\left\langle \taur\right\rangle =1$) or constant Rosseland
optical depth ($\taur=1$), as shown in Fig. \ref{fig:surface_velocity_correlation}.
The thermodynamic properties correlate very well with vertical velocity,
and the vertical velocity is separating the different properties between
the up- and downflows (see SN98). All the thermodynamic properties
exhibit a bimodal distribution due to the inherent asymmetric nature
of the convective energy transport. On the one hand, the stellar plasma
in the upflows has hotter temperatures and lower densities with brighter
intensities located in the granules. On the other hand, the downflows
are composed by cooler temperatures and higher densities with darker
intensities found in the intergranular lane. Furthermore, the (slower)
upflows correlate with higher entropy and ionization, while the (faster)
downflows associate with lower entropy and ionization. In Fig. \ref{fig:surface_velocity_correlation}
we show the mean values of the histograms for $T$, $\rho$ and $\delta I$,
which exhibits typically a s-shape. Moreover, we display the contours
of one-fifth of the maximal probability for the $T$ and $\rho$.
Then one can obtain a temperature jump from the histograms derived
on layers of constant geometrical depth (left panel in Fig. \ref{fig:surface_velocity_correlation}),
where we determined the height of the optical surface on the temporal
average, i.e. $\left\langle \tau_{\mathrm{Ross}}\right\rangle =1$.
At a higher effective temperature the density decreases, while the
velocity rises, thereby leading to an enhancement in the overshooting
of the convective upflows, which is also known as \textquotedbl{}naked
granulation\textquotedbl{} \citep{Nordlund:1990p6720}. In agreement
with the latter, we find that the $T$-jump becomes more distinct
for hotter $\teff$, and the bimodal distribution between the up-
and downflows is more evident (compare the mean values between the
up- and downflows). The upflows have distinctively larger values compared
to averages on layers of constant optical depth (right panel in Fig.
\ref{fig:surface_velocity_correlation}). The velocity correlation
is tighter at layers of constant optical depth ($\tau_{\mathrm{Ross}}=1$),
since the opacity is very $T$-sensitive due to the negative hydrogen
opacity, and layers with similar temperatures are mapped during transformation
to the optical depth (see Paper II). The fluctuations of the upflows
are broader in temperature and narrower for the density (see Fig.
\ref{fig:surface_velocity_correlation}), while the downflows feature
a broad distribution in $\rho$ and smaller ranges in $T$.

\section{Conclusions\label{sec:Conclusions}}

We derived extensive details of stellar granulation by applying the
multiple layer tracking algorithm for the detection of granules imprinted
in the emergent (bolometric) intensity map, which was originally developed
for solar observations. This method works very reliable for different
stellar parameters. Then, we determined for the individual detected
granules properties: diameter, intensity, temperature, density, velocity
and geometry. The granule diameters span a large range, therefore,
we advise the use of a logarithmic equidistant histogram, since otherwise,
the smaller scales are under-resolved, which leads to the misinterpretation
of a large population of small granules. A distinguished dominant
granule size can always be determined with the maximum of the area
contribution function, which is often very close the maximum of the
diameter distribution. Furthermore, we find two distinct fractal dimensions
(slopes of the area-perimeter relation) that are divided at the dominant
granule size. For smaller granules the fractal dimension is always
very close to unity, which points out that these are evenly shaped.
The larger granules have distinctively larger fractal dimensions close
to 2, primarily depending on the effective temperature. For lower
$\teff$ we find fractal dimension being larger. In the case of the
solar simulation, the dual fractal dimensions we find is in contradiction
to the finding by \citet{Bovelet:2001p6138}, who finds only a single
fractal dimension with the same method from solar observations, and
the discrepancy might root in observational constraints. The bifurcation
of the fractal dimension above the dominant granule size arises simply
due to the fragmentation of granules, which will inevitably entail
that the perimeter increases. We studied also the properties prevailing
at the optical surface in our stellar atmosphere simulations. We find
that the corrugation of the optical surface increases for higher $\teff$
and lower $\logg$. Also, we revealed the systematic correlation of
the intensity, temperature and density with the vertical velocity
as a natural consequence of the convective energy transport.

\bibliographystyle{aa}
\bibliography{papers}

\end{document}